\documentclass[aps,prd,twocolumn,showpacs,floatfix,preprintnumbers,amsfont,amsmath,amssymb,nofootinbib, superscriptaddress]{revtex4-1}

\usepackage[utf8]{inputenc}
\usepackage{amsmath}
\usepackage{physics}
\usepackage{comment}
\usepackage{bm}
\usepackage[separate-uncertainty=true]{siunitx}
\usepackage{graphicx}
\usepackage{natbib}
\usepackage[hidelinks]{hyperref}
\usepackage{mathtools}
\usepackage{xcolor}
\usepackage{ctable}
\usepackage[normalem]{ulem}
\usepackage{multirow}
\usepackage[caption=false]{subfig}
\usepackage[export]{adjustbox}

\newcommand{\rmd}{{\rm d}}
\newcommand{\diff}[2]{\frac{\rmd #1}{\rmd #2}}

\newcommand{\rhol}{\rho_{\rm L}}
\newcommand{\ntrig}{{N_{\rm trig}}}
\newcommand{\pastro}{{p_{\rm astro}}}
\newcommand{\pastroi}{{p_{{\rm astro}, i}}}
\newcommand{\pdet}{{p_{\rm det}}}
\newcommand{\chieff}{{\chi_{\rm eff}}}
\newcommand{\chidiff}{{\chi_{\rm diff}}}
\newcommand{\mones}{{m_{1\rm s}}}
\newcommand{\coloneq}{\vcentcolon=}
\newcommand{\ddfrac}[2]{\frac{\displaystyle #1}{\displaystyle #2}}
\newcommand{\lambdainj}{{\lambda'_{\rm inj}}}
\newcommand{\tabline}{\specialrule{.1em}{.05em}{.05em}}
\newcommand{\appropto}{\mathrel{\vcenter{
  \offinterlineskip\halign{\hfil$##$\cr
    \propto\cr\noalign{\kern2pt}\sim\cr\noalign{\kern-2pt}}}}}
\newcounter{modelID}

\newcommand{\model}[1]{\refstepcounter{modelID}\label{model:#1}}
\newcounter{submodelID}

\newcommand{\submodel}[1]{\refstepcounter{submodelID}\label{model:#1}}

\newcommand{\sk}[1]{}

\begin{document}


\title{Binary Black Hole Mergers from LIGO/Virgo O1 and O2: \texorpdfstring{\\}{}
       Population Inference Combining Confident and Marginal Events}

\author{Javier Roulet}
\email{jroulet@princeton.edu}
\affiliation{\mbox{Department of Physics, Princeton University, Princeton, New Jersey 08540, USA}}
\author{Tejaswi Venumadhav}
\affiliation{\mbox{School of Natural Sciences, Institute for Advanced Study, 1 Einstein Drive, Princeton, New Jersey 08540, USA}}
\affiliation{\mbox{Department of Physics, University of California at Santa Barbara, Santa Barbara, California 93106, USA}}
\affiliation{\mbox{International Centre for Theoretical Sciences, Tata Institute of Fundamental Research, Bangalore 560089, India}}
\author{Barak Zackay}
\affiliation{\mbox{School of Natural Sciences, Institute for Advanced Study, 1 Einstein Drive, Princeton, New Jersey 08540, USA}}
\affiliation{\mbox{Department of Particle Physics \& Astrophysics, Weizmann Institute of Science, Rehovot 76100, Israel}}
\author{Liang Dai}
\affiliation{\mbox{School of Natural Sciences, Institute for Advanced Study, 1 Einstein Drive, Princeton, New Jersey 08540, USA}}
\affiliation{\mbox{Department of Physics, University of California, Berkeley, 366 LeConte Hall, Berkeley, California 94720, USA}}
\author{Matias Zaldarriaga}
\affiliation{\mbox{School of Natural Sciences, Institute for Advanced Study, 1 Einstein Drive, Princeton, New Jersey 08540, USA}}

\date{\today}
             
\begin{abstract}
We perform a statistical inference of the astrophysical population of binary black hole (BBH) mergers observed during the first two observing runs of Advanced LIGO and Advanced Virgo, including events reported in the GWTC-1 and IAS catalogs. 
We derive a novel formalism to fully and consistently account for events of arbitrary significance.
We carry out a software injection campaign to obtain a set of mock astrophysical events subject to our selection effects, and use the search background to compute the astrophysical probabilities $\pastro$ of candidate events for several phenomenological models of the BBH population. We emphasize that the values of $\pastro$ depend on both the astrophysical and background models. Finally, we combine the information from individual events to infer the rate, spin, mass, mass-ratio and redshift distributions of the mergers.
The existing population does not discriminate between random spins with a spread in the effective spin parameter, and a small but nonzero fraction of events from tidally-torqued stellar progenitors. 
The mass distribution is consistent with one having a cutoff at $m_{\rm max} = 41^{+10}_{-5}\,\rm M_\odot$, while the mass ratio favors equal masses; the mean mass ratio $\overline q> 0.67$. The rate shows no significant evolution with redshift.
We show that the merger rate restricted to BBHs with a primary mass between 20--\SI{30}{M_\odot}, and a mass ratio $q > 0.5$, and at $z \sim 0.2$, is 1.5--\SI{5.3}{Gpc^{-3} yr^{-1}} (90\% c.l.); these bounds are model independent and a factor of $\sim 3$ tighter than that on the local rate of all BBH mergers, and hence are a robust constraint on all progenitor models.
Including the events in our catalog increases the Fisher information about the BBH population by $\sim 47\%$, and tightens the constraints on population parameters. 
\end{abstract}

\maketitle

\section{Introduction}

The Advanced LIGO \cite{LIGOdetector} and Advanced Virgo \cite{Virgodetector} detectors have detected gravitational waves (GW) from a number of compact binary mergers since the start of the first observing run in 2015. 
Summarizing their first and second observing runs (O1 and O2), the LIGO and Virgo Collaboration (LVC) released a catalog (GWTC-1, see Ref.~\cite{GWTC-1}) with ten BBH mergers and one binary neutron star (BNS) merger. 
The recently concluded third observing run (O3) has yielded a second BNS merger \cite{GW190425} and three new BBH mergers \cite{GW190412, GW190814, GW190521}, with more to be announced. 

The LVC has released the raw strain data from O1 and O2 \citep{GWOSC}, and several independent teams have reanalyzed this dataset \cite{1-OGC, 2-OGC, Antelis2019, pipeline, BBH_O2, fishing}.
In previous work, we identified nine additional BBH events with significance ranging from high to marginal levels~\cite{pipeline, GW151216, BBH_O2, fishing} (for simplicity, we will refer to these events using the abbreviation IAS, after the Institute for Advanced Study). Several of these events were independently confirmed by Ref.~\citep{2-OGC}, who also identified another significant event.

Despite being among the most detectable and accurately modeled GW sources, the origin of merging BBHs remains unclear. A variety of astrophysical formation mechanisms have been proposed, including isolated binary stellar evolution through a common envelope phase \cite{nelemans2001gravitational, belczynski2002comprehensive, voss2003galactic, belczynski2007rarity, belczynski2008compact, dominik2013double, belczynski2014formation, mennekens2014massive, spera2015mass, eldridge2016bpass, stevenson2017formation, mapelli2017cosmic, giacobbo2017merging, mapelli2018cosmic, kruckow2018progenitors, giacobbo2018progenitors}, chemically homogeneous stellar evolution \cite{marchant2016new, de2016chemically, mandel2016merging}, or dynamic capture and hardening of binaries in dense stellar systems such as globular clusters \cite{zwart1999black, o2006binary, sadowski2008total, downing2010compact, downing2011compact, Samsing:2013kua, PhysRevLett.115.051101, rodriguez2016binary, askar2016mocca}, nuclear clusters \cite{antonini2016merging, petrovich2017greatly}, and young open clusters \cite{ziosi2014dynamics, mapelli2016massive, banerjee2017stellar, chatterjee2017dynamical}. Alternatively, mergers can be prompted by interactions with gas and stars in AGN disks \cite{mckernan2012intermediate, stone2016assisted, bartos2017rapid}, or through the Kozai--Lidov effect in the presence of a supermassive black hole \cite{antonini2012secular} or in triple (or higher multiplicity) systems \cite{antonini2014black, kimpson2016hierarchical, antonini2017binary, liu2018black, Hamers2015}.

The growing number of detections has spawned many efforts to statistically characterize the population of these systems, with the main goal of comparing the observed sample statistics with the predictions of different proposed formation channels.
The distributions of the BBH merger rate, masses, spins and redshifts have been studied in the literature \cite{Vitale2017, Talbot2017, LVC_O1, Fishbach2017, Hotokezaka2017, Fishbach2018, Farr2018, Wysocki2018, Wysocki2019, Roulet2019, LVC_pop_O2, Fishbach2020}.
Several of these works were based on the entirety or a subset of the ten confident BBH detections reported in GWTC-1.
Other recent works also included events from the IAS catalog in the BBH population analyses \cite{Piran2020, Gayathri2020, Galaudage2020, Pratten2020}. However, when considering the events as a population, we have to appropriately account for the marginal significance of some of these events: for such events, the probability of astrophysical origin itself can depend on the population model being considered (this was previously noted in Ref.~\cite{2-OGC}). Moreover, the probability of astrophysical origin for a candidate depends on the search pipeline that found it, since the definition involves the levels of comparable foreground and background triggers, subject to the same selection effects. 
Hence, the sensitivity of the search pipeline is a crucial piece of information when inferring astrophysical event rates and correcting for selection effects. In this work, we inject synthetic signals into the O1 and O2 runs to empirically measure the spacetime-volume our pipeline is sensitive to. 

In this paper, we develop a general framework to treat the problem of population inference using detections of arbitrary significance, and apply it in an analysis that accounts for the BBH events in the GWTC-1 and IAS catalogs.
This problem was first studied by \citet{Gaebel2019}, who introduced a formalism for population inference using a mixture of noise and signal triggers. They demonstrated its implementation for a simulation on a simplified parameter space with analytical prescriptions for the foreground and background distributions. Our work expands on this by developing a framework that can be cast in terms of a small number of quantities that are straightforward to compute, and thus more amenable to real-data applications, and implementing it on the O1 and O2 datasets.
\citet{Galaudage2020} developed a different implementation and applied it to include the events in the IAS catalog. We address some issues with this treatment, and how our formalism deals with them, in Appendix~\ref{app:differences}.

Three BBH mergers from the O3 run have been recently reported \cite{GW190412, GW190814, GW190521}. We exclude them from the present analysis since as highlights from a (yet unreleased) O3 catalog, they cannot be simply added to a set of homogeneously selected events for population inference.

We organize the rest of the paper as follows: in \S\ref{sec:framework}, we derive the likelihood of a population model as a function of events of arbitrary significance,  and then we present the algorithm to compute in practice the various quantities involved.
In addition, we estimate the gain in the Fisher information from the inclusion of marginal triggers, and use it as a guide to set a convenient threshold for which triggers to include in the analysis.
In \S\ref{sec:implications}, we report new constraints on the astrophysical population of BBH mergers. In particular, we study the merger rate distribution's dependence on spin, mass, mass ratio and source redshift.
Finally, we draw conclusions in \S\ref{sec:conclusions}.
We quantify the sensitive volume-time of our search pipeline by means of software injection in Appendix~\ref{app:sensitivity}.
We describe technical details of computing the astrophysical probability, $\pastro$, in Appendix~\ref{app:pastro}. 
We address the relation to previous related work in Appendix~\ref{app:differences}. We provide evidence that our method is robust to importance sampling stochastic errors in Appendix~\ref{app:robustness}.

\section{Framework} \label{sec:framework}

In this section we derive the likelihood function for the set of triggers above a given threshold in a pipeline due to a distribution of BBHs, present an algorithm for its practical evaluation, and derive the amount of Fisher information contained in the marginal triggers. The likelihood constrains which population models are consistent with the data.
We will follow the notation of \citet{Mandel2019}, see also Refs.~\cite{Thrane2019, Vitale2020} for an introductory treatment.

\subsection{Model likelihood}

As a preliminary matter, we define the data as the observable quantities that detectors output, along with any quantities derived from this that we use in the search. 
A datum $d$ consists of (a) a measured strain timeseries at each detector, long enough to capture all astrophysical information available in a putative signal, and (b) derived quantities, e.g. detection statistics or statistics used for signal quality tests. 
Note that this excludes BBH parameters such as masses and spins, which are not directly observable.
The full dataset (here, O1 and O2) can be thought of a large set of points in this high-dimensional space, one for every datum.
These realizations contain detector noise plus, in comparatively very few cases, astrophysical signals.

We restrict the analysis to a small set of selected data realizations $\{d_i\}$ (``triggers'') defined under some criteria such that each excluded individual datum is very unlikely to contain an astrophysical signal. This set may contain both secure and marginal events.
We will select the triggers using the search pipeline described in Ref.~\cite{pipeline} and a threshold on its detection statistic (henceforth ``detection threshold'', we discuss our choice in \S\ref{ssec:threshold_choice}); it is the task of the search pipeline to compute the detection statistic for all data.
We will assume triggers to be independent: since triggers are rare, the fact that a trigger uses a sample does not affect significantly the total number of available samples for other triggers to happen. The assumption of independence could be invalid, though, if different triggers of astrophysical origin were produced by multiple images of a single gravitationally lensed source \cite{DaiO2LensingSearch}; we will not consider such possibility in this analysis.

Under this assumption, the search for BBHs in the full dataset is a Poisson process that generates a set $\{d_i\}$ of $\ntrig$ triggers in the above-threshold subregion of the phase space of data.  The likelihood of observing this set of triggers is given by the Poisson distribution
\begin{multline}
    P(\ntrig, \{d_i\} \mid \lambda) \\
    = \frac{e^{-N_a(\lambda) - N_b}}{\ntrig!} \prod_{i = 1}^\ntrig
        \left(\diff{N_a}{d} \bigg|_{d_i} (\lambda)
        + \diff{N_b}{d} \bigg|_{d_i} \right),
        \label{eq:likelihood}
\end{multline}
where $N_a(\lambda)$ is the expected number of triggers in the set with astrophysical origin under a model for the source population described by a set of parameters $\lambda$, $N_b$ is the expected number of noise background triggers in the set, and the terms $\rmd N_a/\rmd d, \rmd N_b/\rmd d$ are the rate densities for triggers under the astrophysical and background hypotheses. 

We express the expected rate density of astrophysical triggers in terms of the physical merger rate through
\begin{equation} \label{eq:dNa/dd}
    \diff{N_a}{d}\bigg|_{d}(\lambda)
    = \int \rmd \theta\, P(d \mid \theta)\,
        \frac{\rmd N_a}{\rmd \theta}(\theta \mid \lambda),
\end{equation}
where $\theta$ are a set of parameters that characterize each merger (e.g. masses, spins, distance, sky position, orbital orientation, time, etc.) and $P(d \mid \theta)$ is the parameter likelihood. For triggers that pass all signal quality tests, a Gaussian noise model is typically a good description of the parameter likelihood:
\begin{equation} \label{eq:Gaussian_like}
    P(d \mid \theta)
    \propto \exp(-\frac 12 \langle d - h(\theta) \mid d - h(\theta) \rangle),
\end{equation}
where $h(\theta)$ is the GW strain model and, in a slight abuse of notation, $d$ is the measured strain. As is standard in the GW literature \cite{Thorne1987}, the argument of the exponential in Eq.~\eqref{eq:Gaussian_like} is the inverse-variance weighted inner product between two real-valued time series $x$ and $y$,
\begin{equation}
    \langle x \mid y \rangle = 4 \Re \int_0^\infty \rmd f\, \frac{\tilde x^\ast(f)\, \tilde y(f)}{S_n(f)},
\end{equation}
 where $S_n(f)$ is the one-sided power spectral density (PSD) of the detector noise, tildes indicate Fourier transforms, and asterisks complex conjugation. 

The expected rate density of background triggers also depends on the detection pipeline. We will estimate it using the method of timeslides, see Appendix~\ref{app:pastro} for further details.

The expected number of astrophysical triggers is
\begin{equation}
    \begin{split} \label{eq:Na}
        N_a(\lambda)
        &= \int_{d>\rm th} \rmd d\, \diff{N_a}{d}\bigg|_{d} (\lambda) \\
        &= \int \rmd \theta\, \diff{N_a}{\theta}(\theta \mid \lambda)\,
            \pdet(\theta).
    \end{split}
\end{equation}
In the first line, the $d$ integral runs over all data realizations that would result in a pipeline statistic above the detection threshold.
In the second line, we introduced the detection efficiency for a source with parameters $\theta$:
\begin{equation}
    \pdet(\theta) = \int_{d > \rm th} \rmd d\, P(d \mid \theta),
\end{equation}

The probability $\pastroi$ that the $i$th trigger is of astrophysical origin depends on models for both astrophysical events and noise triggers. In this work, we will fix the model for noise triggers, but vary the astrophysical one. Given an astrophysical model described by parameters $\lambda$, 
\begin{equation} \label{eq:pastro}
    \pastroi(\lambda)
    = \frac{\rmd N_a(\lambda)}{\rmd N_a(\lambda)+\rmd N_b} \bigg|_{d_i}.
\end{equation}

For practical evaluation, we rescale Eq.~\eqref{eq:likelihood} into the following form while keeping the dependence on the model parameters $\lambda$,
\begin{widetext}
    \begin{equation} \label{eq:like_ratio}
        \begin{split}
            P(\ntrig, \{d_i\} \mid \lambda)
            &\propto \frac{P(\ntrig, \{d_i\} \mid \lambda)}
                          {P(\ntrig, \{d_i\} \mid \lambda_0)} \\
            &\propto e^{- N_a(\lambda)} \prod_{i = 1}^\ntrig
                \frac{\rmd N_a(\lambda) + \rmd N_b}
                     {\rmd N_a(\lambda_0) + \rmd N_b}\bigg|_{d_i} \\
            &= e^{- N_a(\lambda)} \prod_{i = 1}^\ntrig
                \left[\frac{\rmd N_a(\lambda)}{\rmd N_a(\lambda_0)}\bigg|_{d_i}
                \frac{\rmd N_a(\lambda_0)}
                     {\rmd N_a(\lambda_0) + \rmd N_b} \bigg|_{d_i}
                + \frac{\rmd N_b}
                       {\rmd N_a(\lambda_0) + \rmd N_b} \bigg|_{d_i} \right] \\
            &= e^{- N_a(\lambda)} \prod_{i = 1}^\ntrig \left[
                \diff{N_a(\lambda)}{N_a(\lambda_0)}\bigg|_{d_i} \pastroi(\lambda_0)
                + \big(1 - \pastroi(\lambda_0)\big) \right],
        \end{split}
    \end{equation}
\end{widetext}
where $\lambda_0$ corresponds to a fiducial source population model that we are free to choose. Equation \eqref{eq:like_ratio} converges to a meaningful number as one relaxes the detection threshold and includes arbitrarily insignificant triggers with $\pastro \to 0$ (we will discuss this point further in \S\ref{ssec:threshold_choice}). In the opposite limit in which we only include events with absolute certainty of astrophysical origin ($\pastro = 1$), we recover the standard result (e.g. in the notation of \citet{Mandel2019})
\begin{equation} \label{eq:pastro=1}
    \begin{split}
        {}&P(\ntrig, \{d_i\} \mid \lambda, \pastroi=1) \\
        &\quad= \frac{e^{-N_a(\lambda)}}{\ntrig!} \prod_{i=1}^\ntrig
            \diff{N_a}{d}\bigg|_{d_i} (\lambda) \\
        &\quad= \frac{e^{-N_a(\lambda)}}{\ntrig!}
            \left[ N_a(\lambda) \right]^\ntrig
            \prod_{i=1}^\ntrig \ddfrac{\int \rmd \theta\, P(d_i \mid \theta)
                                       \diff{N_a}{\theta} (\theta \mid \lambda)}
            {N_a(\lambda)} \\
        &\quad= P(\ntrig \mid \lambda) \prod_{i=1}^\ntrig \frac
            {\int \rmd \theta\, P_{\rm pop}(\theta \mid \lambda')\,P(d_i \mid \theta)}
            {\int\rmd\theta\, P_{\rm pop}(\theta \mid \lambda')\, \pdet(\theta)},
    \end{split}
\end{equation}
where $P(N_{\rm trig}\mid \lambda)$ follows standard Poisson statistics (this can also be obtained by marginalizing over $\{d_i\}$ in Eq.~\eqref{eq:likelihood}). In the above formula, $\lambda'$ are the population parameters that characterize the shape of the un-normalized astrophysical distribution $P_{\rm pop}$, separated out from an overall merger rate, which we will call $R$. The overall rate $R$ cancels inside the product since both $\rmd N_a/\rmd \theta$ and $N_a$ are linearly proportional to it. However, note that once we include events with $0 < \pastro < 1$, the value of $\pastro(\lambda)$ depends on the rate even at fixed population shape and such a clean separation does not occur.

Nevertheless, we can exploit the linear dependence of $N_a(\lambda)$ and $\rmd N_a/\rmd \theta$ on the rate parameter to evaluate these terms efficiently.
We make explicit the decomposition of the population parameters $\lambda$ into rate $R$ and shape $\lambda'$:
\begin{equation} \label{eq:R*f}
    \frac{\rmd N_a}{\rmd \theta}(\theta \mid \lambda)
    = R\, f(\theta \mid \lambda'),
\end{equation}
where $f(\theta\mid \lambda')$ is normalized according to
\begin{equation}
    \lim_{V\to0}\frac{1}{VT}\, \int_{VT} \rmd\theta\, f(\theta \mid \lambda') = 1
\end{equation}
in the local Universe, so that $R$ is the local merger rate per unit time per unit volume. Note that the source distance and the arrival time of the signal are among the parameters $\theta$, and for these we do not normalize their distribution to integrate to unity over some domain since they do not have a natural scale.
From Eqs.~\eqref{eq:dNa/dd} and \eqref{eq:R*f}, the astrophysical number density ratio for the $i$th trigger in Eq.~\eqref{eq:like_ratio} is
\begin{equation} \label{eq:w_def}
    \diff{N_a(\lambda)}{N_a(\lambda_0)}\bigg|_{d_i}
    = \frac{R}{R_0}\, w_i(\lambda';\lambda'_0),
\end{equation}
where we define
\begin{equation} \label{eq:wi}
    w_i(\lambda';\lambda'_0) \coloneq
        \frac{\int \rmd \theta\, P(d_i \mid \theta)\, f(\theta \mid \lambda')}
        {\int \rmd \theta\, P(d_i \mid \theta)\, f(\theta \mid \lambda_0')}.
\end{equation}
Likewise, the expected detection rate is
\begin{equation} \label{eq:R*g}
    N_a(\lambda) = R\cdot \overline{VT}(\lambda'),
\end{equation}
where, in accord with Eq.~\eqref{eq:Na},
\begin{equation} \label{eq:g}
    \overline{VT}(\lambda') = \int \rmd \theta\, f(\theta \mid \lambda')\, \pdet(\theta)
\end{equation}
is the population-averaged sensitive volume-time of the detector network (analogous to $\alpha(\lambda')$ of \citet{Mandel2019} but with a different normalization choice; equivalent to $\mathcal V(\Lambda)$ of \citet{Galaudage2020}). $\overline{VT}(\lambda')$ depends on the search pipeline and detection threshold used.

In terms of these quantities, Eq.~\eqref{eq:like_ratio} takes the form
\begin{equation} \label{eq:like_ratio2}
    \begin{split}
        &P(\ntrig, \{d_i\} \mid \lambda) \\
        &\qquad\propto e^{- R\,\overline{VT}(\lambda')}\, \prod_{i = 1}^\ntrig \bigg[
            \frac{R}{R_0}\, w_i(\lambda';\lambda'_0)\,
            \pastroi(\lambda_0) \\
        &\qquad\qquad\qquad\qquad\qquad~ + \left( 1 - \pastroi(\lambda_0) \right) \bigg].
    \end{split}
\end{equation}

\subsection{Likelihood evaluation} \label{ssec:likelihood_evaluation}
In order to evaluate Equation~\eqref{eq:like_ratio2}, we need to evaluate three types of terms: $w_i(\lambda';\lambda'_0)$, $\overline {VT}(\lambda')$ and $\pastroi(\lambda_0)$.

We estimate $w_i(\lambda'; \lambda'_0)$ from the integral in Eq.~\eqref{eq:wi} using a Monte Carlo method:
\begin{equation} \label{eq:wi_est}
    w_i(\lambda';\lambda'_0) \approx \mathcal W_i(\lambda';\lambda'_0) \coloneq \ddfrac
        {\sum_{j=1}^{S_i} \frac{1}{\pi(\theta^i_j)}\, f(\theta^i_j \mid \lambda')}
        {\sum_{j=1}^{S_i} \frac{1}{\pi(\theta^i_j)}\, f(\theta^i_j \mid \lambda'_0)},
\end{equation}
where $\{\theta^i_j: j=1,\ldots,S_i\}$ are samples from the posterior distribution of the parameters for the $i$th trigger, obtained under a prior $\pi(\theta)$.

Similarly, we can evaluate $\overline{VT}(\lambda')$ by reweighting a set of injections \cite{Tiwari2018}. We add synthetic signals to the data in software and run the detection pipeline (with vetoes and choices that are as close as possible to those in the `production' run\footnote{We change some choices in order to keep computational cost manageable, as detailed in Appendix \ref{sec:inj_campaign}.}) to determine which injections would have been found. From Eq.~\eqref{eq:g} we can construct the estimator
\begin{equation} \label{eq:g_est}
    \begin{split}
        \overline{VT}(\lambda') \approx \mathcal{\overline{VT}}(\lambda')
        \coloneq \frac{1}{N_{\rm inj}} \sum_{j > \rm th}
            \frac{f(\theta_j \mid \lambda')}
                 {P(\theta_j \mid \lambda'_{\rm inj})}.
    \end{split}
\end{equation}
Here, we denote by $P(\theta \mid \lambda'_{\rm inj})$ the distribution of source parameters from which we generate injections. Note that $N_{\rm inj}$ is the total (found and missed) number of injections, but the sum runs only over those above the detection threshold.

Finally, we have to compute the reference $\pastro(\lambda_0)$ for all triggers under consideration. Note that these correspond to a particular astrophysical model $\lambda_0$, so we cannot use the numbers reported by a pipeline at face value without regard to $\lambda_0$. In \S\ref{ssec:distribution_choice} we describe our choice of $\lambda_0$.
According to Eq.~\eqref{eq:pastro}, $\pastro$ requires estimating both the foreground and the background of the search pipeline. We estimate the foreground using injections and the background using timeslides (see Appendix~\ref{app:pastro} for details). We report the values of $\pastro(\lambda_0)$ in Table~\ref{tab:pastro}.

As long as the population models $\lambda$ of interest do not have too many parameters, we can evaluate the rate-independent estimators $\{\mathcal W_i(\lambda';\lambda'_0)\}$ and $\mathcal{\overline{VT}}(\lambda')$ on an auxiliary $\lambda'$ grid. We then use them to evaluate Eq.~\eqref{eq:like_ratio2} on a $\lambda$ grid, that incorporates the dependence with rate avoiding redundant reevaluation of $\{\mathcal W_i(\lambda';\lambda'_0)\}$ and $\mathcal{\overline{VT}}(\lambda')$. Note that this procedure extends readily to a situation where $\rmd N_a$ depends linearly on multiple population parameters, a commonly encountered case being the branching ratios of a ``mixture model'' which consists of a linear combination of several sub-populations.

The BBH merger GW170608 occurred while one LIGO detector was not in nominal observing mode~\cite{GW170608}. Data from such periods are not publicly accessible so our injections do not simulate this type of events. Although we cannot use GW170608 to inform the astrophysical rate, the event contains valuable information about the shape parameters $\lambda'$. In order to include GW170608 consistently in our analysis, we single it out as an additional event with $\pastro=1$ that is not counted in $N_a(\lambda)$. We replace Eq.~\eqref{eq:like_ratio} by
\begin{multline} \label{eq:add_events}
    P(\{d_i\}, \ntrig \mid \lambda) \\
    = P(\{d_{i\neq\rm GW170608}\}, \ntrig-1 \mid \lambda)\, P(d_{\rm GW170608} \mid \lambda'),
\end{multline}
where we choose to normalize
\begin{equation}
    P(d_{\rm GW170608} \mid \lambda') = \frac 1{N_a(\pastro \approx 1)} \diff{N_a}{d}\bigg|_{\rm GW170608}
\end{equation}
so that it integrates to 1 over all non-observing-mode data realizations that would have yielded a $\pastro \approx 1$ event. We implement Eq.~\eqref{eq:add_events} under the approximation $N_a(\pastro \approx 1) \appropto N_a(\lambda)$, which in practice amounts to dividing Eq.~\eqref{eq:like_ratio} by $N_a(\lambda)$ (and in general, by $[N_a(\lambda)]^n$ to include $n$ additional events).

\subsection{Choice of reference and injection distributions}\label{ssec:distribution_choice}

The Monte Carlo estimators $\{\mathcal W_i(\lambda';\lambda'_0)\}$ and $\mathcal{\overline{VT}}(\lambda')$ (Eqs.~\eqref{eq:wi_est} and \eqref{eq:g_est}) are unbiased, in the sense that their expectation values are
\begin{align}
    \langle \mathcal W_i (\lambda';\lambda'_0) \rangle &= w_i(\lambda';\lambda'_0) \\
    \langle \mathcal{\overline{VT}}(\lambda') \rangle &= \overline{VT}(\lambda')
\end{align}
regardless of the choice of reference and injection populations, $\lambda'_0$ and $\lambdainj$.
However, due to the finite number of samples used, they have variances that depend on the choice of $\lambda'_0$ and $\lambdainj$. In this subsection, we discuss choices that allow robust estimation of $w_i(\lambda';\lambda'_0)$ and $\overline{VT}(\lambda')$.

The main requirement for these importance sampling estimators is that the proposal distribution from which the samples are taken does not vanish at places where the integrand (target distribution) is nonzero, lest the reweighting become pathological. Thus, neither the parameter estimation prior $\pi(\theta)$ nor the reference population $f(\theta \mid \lambda'_0)$ in Eq.~\eqref{eq:wi_est} should vanish anywhere the likelihood $P(d_i \mid \theta)$ has support. In general, the estimator variance will be smaller when the proposal distribution more closely matches the target distribution.

The effective spin
\begin{equation}
    \label{eq:chieff}
    \chieff \coloneq \frac{\chi_{1z} + q \chi_{2z}}{1 + q}
\end{equation}
is the spin variable that can be measured best, where $q = m_2 / m_1$ is the mass ratio and $\chi_{1z}, \chi_{2z}$ are the dimensionless spin components in the direction of the binary's orbital angular momentum. 
Spin components in the orbital plane are harder to measure than and no evidence for or against them has been found for any of the mergers of O1 and O2. The same is true for gravitational radiation modes beyond the quadrupolar $(\ell, \abs{m}) = (2, 2)$. To simplify the analysis, we will neglect in-plane spin components and higher-order modes in the following, and we do not expect that this will change the results significantly. In particular, we will use the aligned-spin, quadrupolar radiation waveform approximant \texttt{IMRPhenomD} \cite{Khan2016} to model gravitational wave signals throughout this work. For parameter estimation, we use the relative binning method for likelihood evaluation \cite{Zackay2018} coupled to the \texttt{PyMultinest} sampler \cite{Buchner2014}.

We adopt a parameter estimation prior $\pi(\theta)$ that is uniform in detector-frame masses, effective spin $\chieff$ and luminosity volume. 
Some events, most notably GW151216 \cite{GW151216}, have high effective spins, so a prior that does not vanish for extreme values of $\chieff$ is convenient.
We parametrize the two spins in terms of the well measured $\chieff$ and a poorly measured variable $\chidiff \coloneq (q \chi_{1z} - \chi_{2z}) / (1+q)$ that controls how much either binary component contributes to $\chieff$. We implement the spin prior as flat in $\chieff$, and flat in $\chidiff$ conditioned on $\chieff$ within the Kerr bound $\abs{\chi_{1,2}} < 1$.

For the reference population $f(\theta \mid \lambda'_0)$, we narrow our focus to $\theta = \{\mones, q, \chieff, D_L\}$, which are the measurable variables that lack a natural prior, and adopt the following (factorized) joint distribution: truncated power-law in the primary source-frame mass $\mones$ and uniformity in $q$, $\chieff$ and luminosity volume, with $\lambda'_0$ that lie in the bulk of reported posterior distributions \cite[see e.g.][]{LVC_pop_O2}:
\begin{equation} \label{eq:f0}
    f(\mones, q, \chieff, D_L \mid \lambda'_0) \propto m_{1 \rm s}^{-\alpha_0} D_L^2
\end{equation}
for $m_{1\rm min} < \mones < m_{1 \rm max}$ and $q_{\rm min} < q < 1$, with $\alpha_0 = 2.35$, $m_{1 \rm min} = \SI{3}{M_\odot}$, $m_{1 \rm max} = \SI{120}{M_\odot}$ and $q_{\rm min} = 1/20$. These ranges are broad enough to encompass the likelihood support of all the triggers we include. Throughout this work we will implicitly assume a uniform distribution for $\chidiff$, arrival time, orbital phase, orbital orientation and sky position. When computing $\pastro(\lambda_0)$ we will use a fiducial rate $R_0 = 10^{1.5}\,{\rm Gpc}^{-3}\,{\rm yr}^{-1}$. Note that the choice of $\lambda_0$ does not affect the final results (provided the reweighing process presents no pathologies, as we demonstrate in Appendix~\ref{app:robustness}) and therefore using previous analyses to inform our choice does not bias our conclusions.

Finally, we choose an injection distribution (Eq.~\eqref{eq:g_est}) that approximately matches the integrand in Eq.~\eqref{eq:g}. We adopt
\begin{align}
    P(\theta \mid \lambdainj)
    &= Z^{-1}\, f(\theta \mid \lambda'_0)\, \hat p_{\rm det}(\theta), \label{eq:pinj} \\
    \begin{split}
        Z &= \int \rmd \theta\, f(\theta \mid \lambda_0')\, \hat p_{\rm det}(\theta) \\\label{eq:Z}
        &\approx \SI{7.1}{Gpc^3}\,T_{\rm obs},
    \end{split}
\end{align}
where $\hat p_{\rm det}(\theta)$ is some semianalytical approximation of $\pdet(\theta)$, $T_{\rm obs}$ is the observation time span on which the injections are made, and the value of $Z$ reported in Eq.~\eqref{eq:Z} corresponds to the choices that follow. We use
\begin{equation} \label{eq:p_fid}
    \hat p_{\rm det}(\theta) = \int_{\rho_{\rm th}^2}^\infty \rmd \rho^2 \chi^2(\rho^2, 10, \rho_\ast^2(\theta)).
\end{equation}
Here, we use $\rho_{\rm th}^2=60$ as an approximate detection threshold. We define the expected signal-to-noise ratio (SNR) of an event with parameters $\theta$:
\begin{equation}
    \rho_\ast(\theta) = \rho_{1\rm Mpc}(m_1^{\rm det}, q, \chi_{1z}, \chi_{2z})\, A(\alpha, \delta, \iota, \psi, t)\, \frac{\SI{1}{Mpc}}{D_L}.
\end{equation}
This is computed for a fiducial noise PSD, which we define in each frequency bin as the 10th percentile of 200 random \SI{4096}{\second} segments of Hanford and Livingston O2 data. 
In addition, $A = \sqrt{A_{\rm H}^2 + A_{\rm L}^2}$ is the Hanford--Livingston antenna pattern, where the angular factors $0 < A_{\rm H, L} < 1$ can be found, e.g., in Ref.~\cite{Sathyaprakash2009}. 
We define $\rho_{1\rm Mpc} = \langle h \mid h \rangle^{1/2}$ to be the single-detector SNR of an optimally oriented source at a fiducial distance of \SI{1}{Mpc}, which we interpolate on a grid of intrinsic parameters. 

The non-central chi-squared distribution in Eq.~\eqref{eq:p_fid} models the distribution of SNR$^2$ recovered by a search pipeline for a signal with parameters $\theta$ in the presence of Gaussian noise and maximized over ten degrees of freedom \cite{Jaranowski2012}.
Six degrees of freedom model maximization by the pipeline over the signal amplitude, phase and arrival time independently in the two LIGO detectors. The remaining four model maximization over template parameters (our waveform templates are elements of a metric space of up to 4 dimensions \cite{templatebank_paper}). Eq.~\eqref{eq:p_fid} neglects a variety of effects present in the search process, such as template bank discreteness and boundaries, signal coherence across detectors, detector sensitivity variations, noise non-stationarity and non-Gaussianity, and signal quality vetoes. This is acceptable, since $\hat p_{\rm det}$ is only used for choosing a convenient injection distribution: all these effects are accounted for by the injections as per Eq.~\eqref{eq:g_est}. Moreover, they make $\hat p_{\rm det}$ a somewhat optimistic estimate of $p_{\rm det}$, which is desirable as it makes the proposal distribution (Eq.~\eqref{eq:pinj}) broader than the target distribution (Eq.~\eqref{eq:g}).

We generate the set of source parameters $\{\theta_j\}$ for injected signals by drawing samples from the distribution in Eq.~\eqref{eq:pinj} with the \texttt{PyMultinest} sampler \cite{Buchner2014}, with which we simultaneously evaluate the normalization constant $Z$ as reported in Eq.~\eqref{eq:Z}.
Eqs.~\eqref{eq:g_est} and \eqref{eq:pinj} yield
\begin{equation}
    \mathcal{\overline{VT}}(\lambda') = \frac{Z}{N_{\rm inj}} \sum_{j > \rm th} \frac{f(\theta_j \mid \lambda')}{f (\theta_j \mid \lambda_0')\, \hat p_{\rm det}(\theta_j)}.
\end{equation}
We report technical details and results from the injection campaigns in Appendix~\ref{app:sensitivity}.

\subsection{Choice of detection threshold}
\label{ssec:threshold_choice}

We now derive a criterion for the choice of the threshold at which a trigger is deemed sufficiently informative to be included in the analysis.
To that end we compute the Fisher information as a function of detection threshold.
This also serves as an estimate of the amount of information about the BBH population gained by including marginal events in the present analysis.

The information that the data carry about the BBH distribution is encoded in the Fisher matrix
\begin{equation} \label{eq:fisher_def}
    I(\lambda)_{m n} = - \left\langle 
        \partial^2_{m n} \log P(\{d_i\}, \ntrig \mid \lambda)
        \right\rangle_{\{d_i\}, \ntrig}.
\end{equation}
Here, $\partial^2_{mn} = \partial^2/\partial\lambda_m \partial\lambda_n$ is the second derivative with respect to population parameters $\lambda_m$ and $\lambda_n$, and the subscript denotes that the expectation value is over the distribution of observations (Eq.~\eqref{eq:likelihood}):
\begin{equation}
    P(\{d_i\}, \ntrig \mid \lambda)
        = P(\ntrig \mid \lambda) \prod_{i=1}^\ntrig P(d_i \mid \lambda),
\end{equation}
with
\begin{align}
    P(\ntrig \mid \lambda) &= \frac{e^{-N(\lambda)} [N(\lambda)]^\ntrig}{\ntrig!}, \label{eq:P(Ntrig|lambda)}\\
    P(d_i \mid \lambda) &= \frac 1{N(\lambda)} \diff{N(\lambda)}{d} \bigg|_{d_i},
        \label{eq:P(d|lambda)} \\
    N(\lambda) &= N_a(\lambda) + N_b. \label{eq:N}
\end{align}
$N$ is the expected number of astrophysical and background triggers above detection threshold, and hence depends on the threshold choice. Below we quantify how this choice affects $I(\lambda)_{mn}$.
Eqs.~\eqref{eq:fisher_def}, \eqref{eq:likelihood} and \eqref{eq:N} yield
\begin{equation}
    \begin{split}
    \label{eq:fisher_1}
        I(\lambda)_{mn}
        &= -\left\langle -\partial^2_{mn} N + \sum_{i=1}^\ntrig \partial^2_{mn} \log \diff Nd \bigg|_{d_i} \right\rangle_{\{d_i\}, \ntrig} \\
        &= \partial^2_{mn} N
            - \left\langle \sum_{i=1}^\ntrig \frac{\partial^2_{mn} \rmd N / \rmd d}{\rmd N / \rmd d} \right\rangle
            \\
            &\quad+ \left\langle \sum_{i=1}^\ntrig \left(\partial_m \log \diff Nd \right)\left(\partial_n \log \diff Nd \right) \right\rangle.
    \end{split}
\end{equation}
Using Eqs.~\eqref{eq:P(Ntrig|lambda)} and \eqref{eq:P(d|lambda)} we evaluate the first sum:
\begin{equation}
    \begin{split} \label{eq:first_term}
        \left\langle \sum_{i=1}^\ntrig \frac{\partial^2_{mn} \rmd N / \rmd d}{\rmd N / \rmd d} \right\rangle
        &= \langle \ntrig \rangle \int \rmd d \frac 1N \diff Nd \frac{\partial^2_{mn} \rmd N / \rmd d}{\rmd N / \rmd d} \\
        &= \partial^2_{mn} \int \rmd d \diff N d \\
        &= \partial^2_{mn} N.
    \end{split}
\end{equation}
Further, from Eqs.~\eqref{eq:N} and \eqref{eq:pastro} we obtain
\begin{equation}
    \begin{split} \label{eq:second_term}
        \partial_m \log \diff Nd
        &= \diff{N_a}{N}\, \partial_m \log \diff{N_a}{d} \\
        &= \pastro(d) \,\partial_m \log \diff{N_a}{d},
    \end{split}
\end{equation}
which we use to evaluate the second sum in Eq.~\eqref{eq:fisher_1}.
Equations~\eqref{eq:fisher_1}, \eqref{eq:first_term} and \eqref{eq:second_term} yield
\begin{equation} \label{eq:fisher_2}
        I(\lambda)_{mn} = \left\langle \sum_{i=1}^\ntrig p_{{\rm astro}, i}^2 \left(\partial_m \log \diff{N_a}d \right)
        \left(\partial_n \log \diff{N_a}d \right)
        \right\rangle.
\end{equation}
The special case where $\lambda_m = \lambda_n = R$ is the astrophysical merger rate is particularly simple, because $\rmd N_a / \rmd d$ is proportional to the rate:
\begin{equation}
    \frac{\partial}{\partial R} \log \diff {N_a} d = \frac 1 R,
\end{equation}
so the Fisher information about the rate is
\begin{equation} \label{eq:I(R)}
    I(R) = \frac 1{R^2} \left\langle \sum_{i=1}^\ntrig {p_{{\rm astro}, i}^2} \right\rangle.
\end{equation}
In general, the information each trigger carries about the population depends on its parameters and the population model through the terms in parentheses in Eq.~\eqref{eq:fisher_2}, weighted by its $p_{\rm astro}^2$. The threshold should therefore be set such that the summed $p_{\rm astro}^2$ of excluded triggers is much smaller than that of included triggers, while keeping their number manageable.

\begin{figure}
    \centering
    \includegraphics[width=\linewidth]{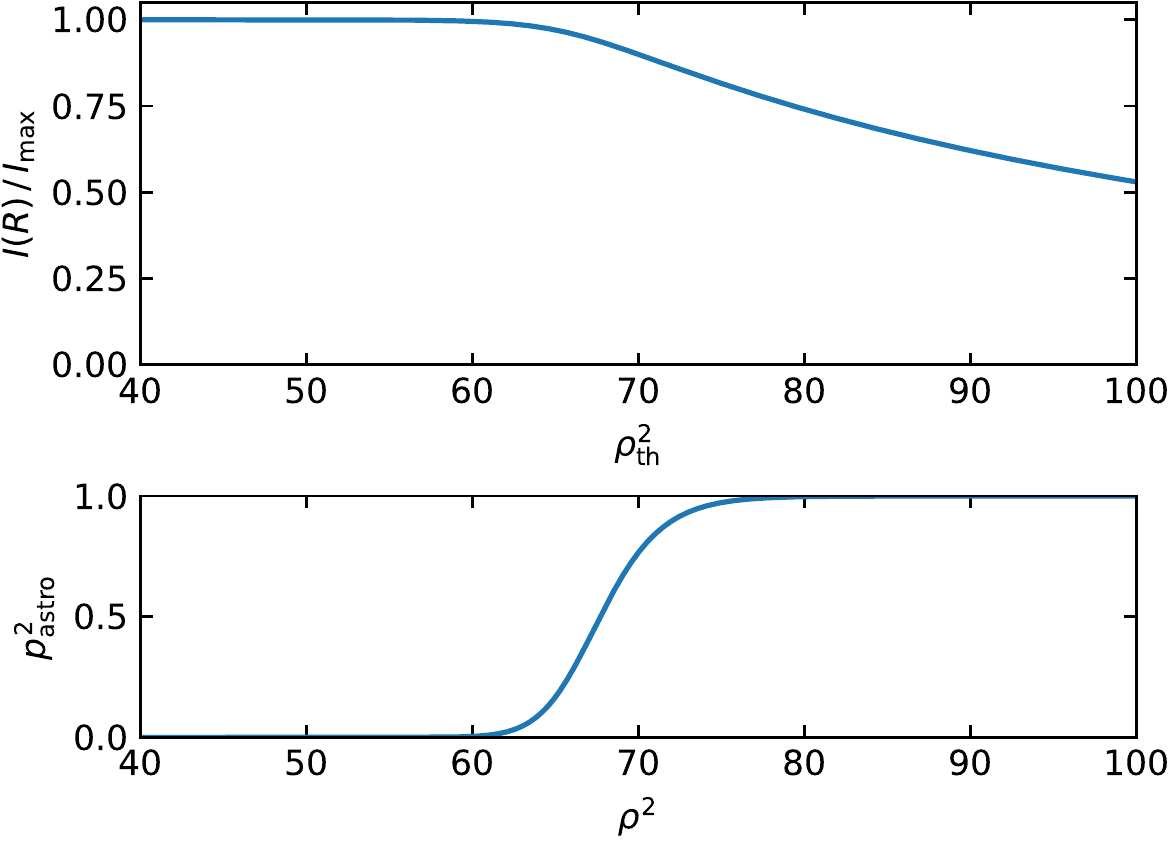}
    \caption{\textit{Top panel:} Fisher information about rate as a function of detection threshold, in a toy model where there is a foreground power-law in SNR with unknown rate and a Gaussian background. The information saturates as the threshold is lowered, thus, faint events carry a limited amount of information.\textit{ Bottom panel:} $p_{\rm astro}^2$ as a function of squared SNR in this model. This quantity determines the average information contributed by each trigger.}
    \label{fig:fisher}
\end{figure}

As an illustrative example, in Fig.~\ref{fig:fisher} we study a simple toy model in which the only parameter measured on triggers is the SNR $\rho$, and there is a population of signals with a power-law distribution $\rmd N_a/\rmd \rho^2 \propto \rho^{-5}$ in a Gaussian background $\rmd N_b / \rmd \rho^2 \propto \exp(-\rho^2/2)$, intended to qualitatively describe features of GW signals \cite{Schutz2011} and detector noise.
We consider the problem of inferring the astrophysical rate $R$, that is, the normalization of the power-law component in this model.
For Fig.~\ref{fig:fisher} we adopt fiducial normalizations so that there are respectively $N_a(\rho^2 > 65) = 15$ and $N_b(\rho^2>65)=1$ expected foreground and background events with a $\rho^2$ louder than 65, roughly matching the numbers observed in O1 and O2. 
The ratio between the foreground and background distributions determines $\pastro(\rho^2)$ through Eq.~\eqref{eq:pastro}, which in turn allows us to compute the Fisher information $I(R)$ with Eq.~\eqref{eq:I(R)}.
In the top panel of Fig.~\ref{fig:fisher} we show the Fisher information as a function of a detection threshold $\rho^2_{\rm th}$ above which triggers are included in the likelihood, Eq.~\eqref{eq:like_ratio}. We find that the information contained in the faint triggers is limited: even though there are many faint signals, their contribution to $I(R)$ is strongly suppressed by $p_{\rm astro}^2$, shown in the bottom panel.
The relative contributions to the information from loud and faint events can be different for other parameters, e.g. if the logarithmic terms in Eq.~\eqref{eq:fisher_2} preferentially select faint events. 
An example of this situation was demonstrated in \citet{Smith2020}, who studied a putative cutoff in the distance distribution.

We choose the detection threshold (for actual triggers as well as for injections) as follows. We include all triggers from O1 and O2 that were found in Hanford--Livingston coincidence with a false-alarm rate (FAR)---within their chirp-mass bank, note that we searched in 5 such banks \cite{pipeline, templatebank_paper}---below one per three times the respective observing run.
With this threshold choice, the summed $p_{\rm astro}^2(\lambda_0)$ of the 30 excluded triggers with lowest FAR is only 0.35, while for the events in Table~\ref{tab:pastro} is 14.65.
In order to include GW170817A consistently, we also include Livingston single-detector triggers from O2 that satisfy the cuts in templates and $\rhol^2$ described in Appendix~\ref{app:pastro} \cite{fishing}. The exception is GWC170402, a Livingston single-detector trigger for which we do not have a satisfactory astrophysical model \cite{fishing}. We exclude GWC170402 from this analysis.
From Table~\ref{tab:pastro} we see that, with some dependence on the population model, including the events from the IAS catalog increases the accumulated $p_{\rm astro}^2$ by $\sim 47\%$, so as a crude estimation we might a priori expect uncertainties in the population parameters to shrink with a factor of order $\sim 1.47^{-1/2} = 0.82$ owing to this additional source of information.

Note that Eq.~\eqref{eq:I(R)} defines the Jeffreys prior for the astrophysical merger rate, $\pi(R \mid \lambda') \propto \sqrt{I(R)}$. In practice, $\pastro$ depends on the rate only for a few near-threshold events and is otherwise very close to either 0 or 1, so we can approximate Eq.~\eqref{eq:I(R)} by
\begin{equation}
    I(R) \approx \frac{N_a(\pastro \approx 1)}{R^2} \appropto \frac{N_a(\lambda)}{R^2} = \frac{\overline{VT}(\lambda')}{R}.
\end{equation}
Under this approximation we find
\begin{equation} \label{eq:jeffreys}
    \pi(R \mid \lambda') \propto \sqrt{\overline{VT}(\lambda') / R},
\end{equation}
or $\pi(N_a) \propto 1/\sqrt{N_a}$, the Jeffreys prior for a single Poisson component. This differs from several other studies that adopt another prior, $\pi(R) \propto 1/\sqrt{R}$. In practice the difference is mild.

\section{Astrophysical implications}
\label{sec:implications}

\begin{table}
\begin{tabular}{lrrrrrrrrr}
\tabline
{} & {$\pastro$} & \multicolumn{8}{c}{$\overline \pastro$}\\
{} &  $\lambda_0$ & Def & \ref{model:gaussian_chieff} & \ref{model:locked} & \ref{model:partially_locked} & \ref{model:m1_powerlaw} & \ref{model:q_powerlaw} & \ref{model:z_powerlaw} & Com \\
\tabline
GW150914   & 1.00 & 1.00 & 1.00 & 1.00 & 1.00 & 1.00 & 1.00 & 1.00 & 1.00\\
GW170809   & 1.00 & 1.00 & 1.00 & 1.00 & 1.00 & 1.00 & 1.00 & 1.00 & 1.00\\
GW170104   & 1.00 & 1.00 & 1.00 & 1.00 & 1.00 & 1.00 & 1.00 & 1.00 & 1.00\\
GW170814   & 1.00 & 1.00 & 1.00 & 1.00 & 1.00 & 1.00 & 1.00 & 1.00 & 1.00\\
GW170729   & 1.00 & 1.00 & 1.00 & 1.00 & 1.00 & 1.00 & 1.00 & 1.00 & 1.00\\
GW170608   & 1.00 & 1.00 & 1.00 & 1.00 & 1.00 & 1.00 & 1.00 & 1.00 & 1.00\\
GW170823   & 1.00 & 1.00 & 1.00 & 1.00 & 1.00 & 1.00 & 1.00 & 1.00 & 1.00\\
GW151226   & 1.00 & 1.00 & 1.00 & 1.00 & 1.00 & 1.00 & 1.00 & 1.00 & 1.00\\
GW151012   & 1.00 & 1.00 & 1.00 & 1.00 & 1.00 & 1.00 & 1.00 & 1.00 & 1.00\\
GW170818   & 0.92 & 0.96 & 0.99 & 0.99 & 0.99 & 0.97 & 0.97 & 0.95 & 1.00\\
\hline
GW170304   & 1.00 & 0.99 & 1.00 & 1.00 & 1.00 & 0.99 & 1.00 & 0.99 & 1.00\\
GW170727   & 0.99 & 0.98 & 0.99 & 0.99 & 0.99 & 0.99 & 0.99 & 0.98 & 1.00\\
GW170121   & 0.98 & 0.99 & 0.98 & 0.98 & 0.97 & 0.99 & 0.99 & 0.98 & 0.99\\
GW170817A  & 0.75 & 0.27 & 0.14 & 0.20 & 0.20 & 0.07 & 0.43 & 0.30 & 0.01\\
GW170202   & 0.62 & 0.69 & 0.78 & 0.75 & 0.73 & 0.72 & 0.68 & 0.68 & 0.76\\
GW170403   & 0.62 & 0.53 & 0.16 & 0.12 & 0.08 & 0.50 & 0.61 & 0.53 & 0.11\\
GW170425   & 0.61 & 0.46 & 0.71 & 0.71 & 0.71 & 0.51 & 0.52 & 0.47 & 0.84\\
GW151216   & 0.51 & 0.47 & 0.00 & 0.01 & 0.00 & 0.55 & 0.52 & 0.47 & 0.00\\
170412B\footnote{This is a new candidate we had previously missed due to an execution error that affected $\approx 0.2\%$ of O2 data, and is not to be confused with 170412 \cite{GWTC-1}. Its detector-frame chirp-mass is $\mathcal M \approx \SI{4.59}{M_\odot}$.} & 0.02 & 0.06 & 0.11 & 0.08 & 0.06 & 0.00 & 0.04 & 0.06 & 0.00\\
\tabline
\end{tabular}

\caption{Value of $\pastro$ for the BBH events considered in this work under various astrophysical models. Events first reported in the GWTC-1 and IAS catalogs are respectively above and below the horizontal line. $\pastro(\lambda_0)$ is the probability of astrophysical origin under the fiducial (unimportant) astrophysical model $\lambda_0$ described in \S\ref{ssec:distribution_choice}. The remaining columns report the marginalized $\overline\pastro$ under the population models considered in \S\ref{sec:implications}: the Default model `Def' (Eq.~\eqref{eq:fhat}) serves as a baseline from which Models A--E explore various departures as follows. \ref{model:gaussian_chieff}: Gaussian in $\chieff$; \ref{model:locked}: tidally-locked progenitors with highly spinning remnants, or \ref{model:partially_locked}: with moderately spinning remnants; \ref{model:m1_powerlaw}: truncated power-law in the primary mass; \ref{model:q_powerlaw}: power-law in the mass ratio; \ref{model:z_powerlaw}: power-law in the redshift. The Combined model `Com' combines the maximum likelihood solutions of models \ref{model:gaussian_chieff}, \ref{model:m1_powerlaw} and \ref{model:q_powerlaw} (\S\ref{ssec:discussion}). $\pastro(\lambda_0)$ approximately matches previously reported values from our pipeline \cite{pipeline, BBH_O2, fishing}, except for GW170818 due to an improvement in the search algorithm, see Appendices \ref{app:sensitivity} and \ref{app:pastro}.
We do not reproduce the results of \cite[table II]{Galaudage2020}, see Appendix~\ref{app:differences} for a discussion.}
\label{tab:pastro}
\end{table}

In this section we report our results on the constraints on the BBH population parameters under various astrophysical models and compare their performances. 
To better visualize the effect of including the events from the IAS catalog, we repeat the analysis with a higher detection threshold (inverse false-alarm rate $\rm IFAR > 3000$ observing runs in our pipeline) that restricts to events in the GWTC-1 catalog \cite{GWTC-1}. GW170608 and GW170818 do not satisfy that cut so we include them ad hoc with $\pastro = 1$, appropriately modifying the likelihood using Eq.~\eqref{eq:add_events} with both events.

We will explore a number of phenomenological models that probe the various measurable source parameters $\mones, q, \chieff, D_L$.
For convenience, we define a ``default'' distribution $\hat f$ that takes the following factorized form:
\begin{equation} \label{eq:fhat}
    \hat f(\mones, q, \chieff, D_L)
        = \hat f_\mones(\mones)\, \hat f_q(q)\, \hat f_\chieff(\chieff)\, \hat f_{D_L}(D_L),
\end{equation}
with
\begin{align}
    \hat f_\mones &\propto m_{\rm 1s}^{-2.35}, \qquad \SI{5}{M_\odot} < \mones < \SI{50}{M_\odot} \label{eq:fm1}\\
    \hat f_q &= \rm U(1/20, 1) \\
    \hat f_\chieff &= \rm U(-1, 1) \\
    \hat f_{D_L} &= \frac{4 \pi D_L^2}{(1+z)^4} \left(1-\frac{D_L}{1+z}\diff{z}{D_L}\right), \label{eq:fD}
\end{align}
and explore the effect of varying individual factors. ${\rm U}(a, b)$ denotes a uniform distribution between $a$ and $b$. Note that $\hat f$ differs from $f(\lambda_0')$ (Eq.~\eqref{eq:f0}) in that it has tighter lower and upper mass cutoffs, the merger rate is a free parameter, and the rate is set to be uniform in comoving volume-time (rather than luminosity volume and observer time) through the factor
\begin{equation}
    \begin{split}
        \diff{t_c}{t}\diff{V_c}{V_L}
        &= \diff{t_c}{t} \frac{D_c^2}{D_L^2} \diff{D_c}{D_L} \\
        &= \frac 1{1+z} \cdot \frac{1}{(1+z)^2} \cdot
            \frac{1}{1+z}\left(1-\frac{D_L}{1+z}\diff{z}{D_L}\right)
    \end{split}
\end{equation}
in Eq.~\eqref{eq:fD}. $f(\lambda_0')$ is chosen to have support throughout all the sensitive parameter space, while $\hat f$ is meant to be a convenient reference point in the space of relevant population models and more closely comparable with the models explored in \cite{LVC_pop_O2}.

To obtain a posterior distribution for the population parameters, the likelihood Eq.~\eqref{eq:like_ratio} has to be multiplied by a prior. We take the priors to be flat except for the rate parameter, where we adopt a Jeffreys prior, Eq.~\eqref{eq:jeffreys}.

For each of the models that we will consider below, in Table~\ref{tab:pastro} we report for every event the astrophysical probability $\overline{\pastroi}$ marginalized over uncertainties in the population model parameters:
\begin{equation} \label{eq:pastro_marg}
    \overline \pastroi = \int \rmd \lambda \, P(\lambda \mid \{d\})\, \pastroi(\lambda).
\end{equation}
To evaluate Eq.~\eqref{eq:pastro_marg}, we obtain $\pastroi(\lambda)$ from Eqs.~\eqref{eq:pastro} and \eqref{eq:w_def} as
\begin{equation}
    \pastroi(\lambda) = \ddfrac{\frac{R}{R_0} w_i(\lambda'; \lambda_0') \, \pastroi(\lambda_0)}{1 + \left( \frac{R}{R_0} w_i(\lambda'; \lambda_0')-1\right) \pastroi(\lambda_0)}
\end{equation}
and marginalize over $\lambda$ by quadrature.

\model{spin}
\subsection{Spin distribution} \label{ssec:spin}
\submodel{gaussian_chieff}
\subsubsection{Model~\ref{model:gaussian_chieff}: Gaussian \texorpdfstring{$\chieff$}{effective spin} distribution}
\label{ssec:gaussian}
We first consider a distribution that is Gaussian in $\chieff$
\begin{equation} \label{eq:f_chieff}
    f_\chieff(\chieff \mid \overline \chieff, \sigma_\chieff) = \mathcal N_{\rm t}(\chieff \mid \overline \chieff, \sigma_\chieff)
\end{equation}
and follows the default $\hat f$ \eqref{eq:fhat} in the other parameters. $\mathcal N_{\rm t}(x \mid \mu, \sigma)$ denotes the normal distribution with mean $\mu$ and dispersion $\sigma$ truncated at $\pm 1$. With this simple model we can explore the symmetry of the $\chieff$ distribution, i.e. whether there is a tendency for alignment between the spins and orbit or not. Dynamical formation models predict the spins of the black holes to be randomly oriented, thus symmetrically distributed about $\chieff=0$, while in the isolated binary evolution scenario a tendency for alignment might be expected (note that while spin-dependent selection effects can bias the detected distribution towards positive effective spins \cite{Ng2018, Roulet2019}, $f$ describes the underlying astrophysical distribution). This model also probes the width of the distribution, which is especially interesting in light of predictions that natal BH spins might be very small, barring tidal torques to the stellar progenitors \cite{Fuller2019}.

\begin{figure}
    \centering
    \includegraphics[width=\linewidth]{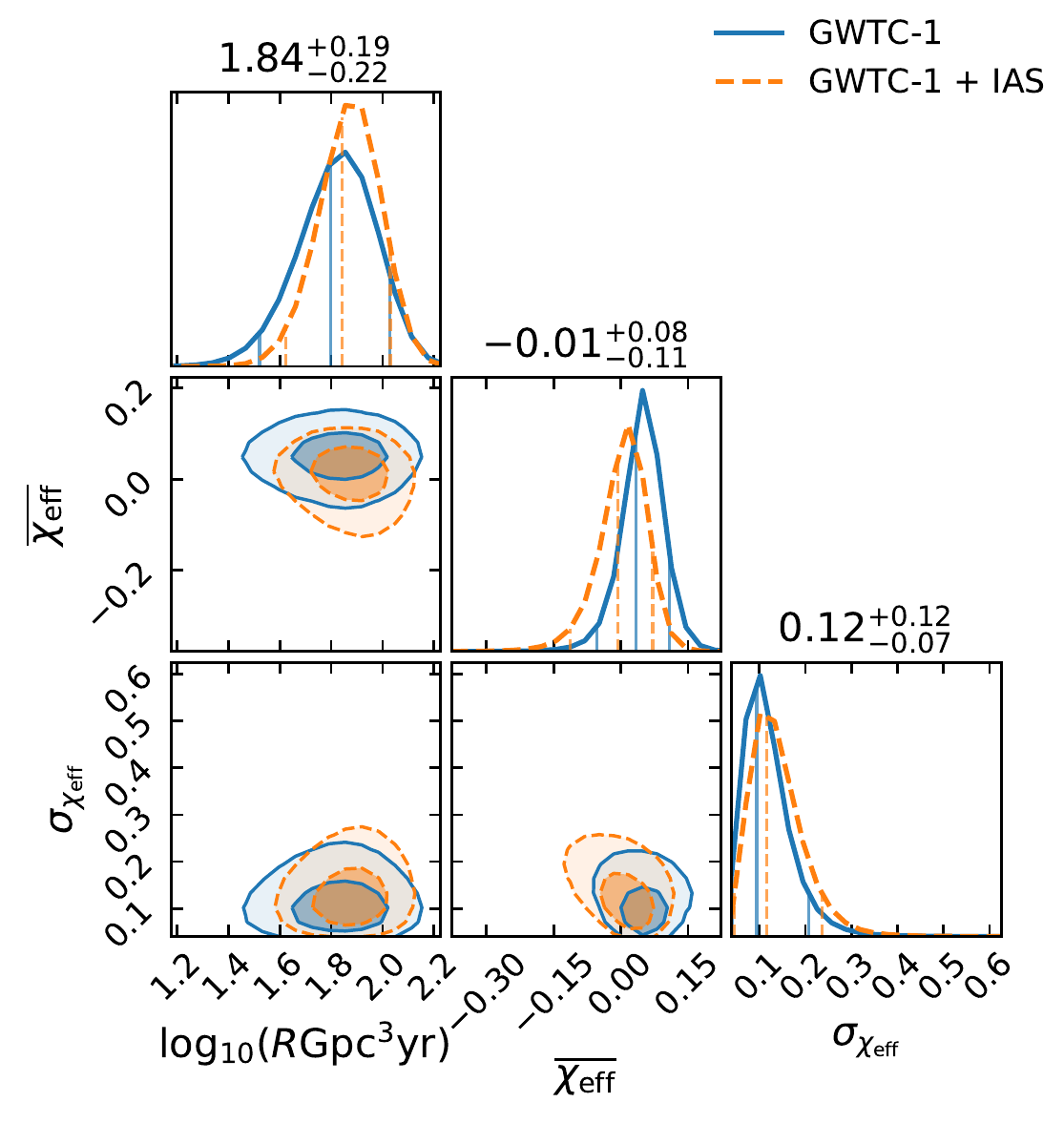}
    \caption{Model \ref{model:gaussian_chieff}: Gaussian $\chieff$ distribution. Solid blue curves show the constraints on the model derived from only the GWTC-1 catalog. Dashed orange curves show the constraints derived from the GWTC-1 and IAS catalogs combined. Two-dimensional contours enclose 50\% and 90\% of the distribution. Vertical lines show the median and 90\% symmetric interval of the one-dimensional posteriors, also reported numerically for the GWTC-1 + IAS analysis.}
    \label{fig:gaussian_chieff}
\end{figure}

The results are shown in Fig.~\ref{fig:gaussian_chieff}. In agreement with previous studies \cite{Farr2018, Wysocki2018, Wysocki2019, Roulet2019, LVC_pop_O2, Miller2020}, the distribution is consistent with $\overline \chieff = 0$ and shows no statistically significant preference for positive $\chieff$, which remains consistent with dynamical formation scenarios.
The width of the distribution is measured to be $\sigma_\chieff \approx 0.13^{+0.12}_{-0.07}$ (median and 90\% c.l.), disfavoring values close to 0 \cite[cf.][]{Miller2020}. Including the events from the IAS catalog yields consistent results. The constraints become tighter for the rate, and broader for the $\chieff$ mean and dispersion. We verified that GW170121, a confident detection with support for negative $\chieff$ \cite{BBH_O2}, rules out the end of the distribution compatible with GWTC-1 with higher $\overline{\chieff}$ and smaller $\sigma_\chieff$, which may drive the change in these constraints. In the rest of this section, we will explore whether the spread of the $\chieff$ distribution can be explained by tidal torques.

\submodel{locked}
\subsubsection{Model~\ref{model:locked}: tidally-locked stellar progenitors}
We now study a model that considers the effects of tides in the BBH progenitor system. 
For field binaries, the typical aftermath of the common-envelope phase is a black hole in a tight orbit with a stripped star. Depending on the binary separation, the star may be subject to strong tides. If the separation is small, corresponding to merger times $\lesssim \SI{e8}{yr}$, the timescale for tidal locking is shorter than the star lifetime and thus it can tidally lock to the orbit. In this case the second-formed black hole would have a high, aligned spin. The tidal-locking timescale depends strongly on the separation, so that for greater separations tides quickly become negligible. In addition, the maximum separation allowed for a circular binary to merge within the age of the Universe is comparable to this distance scale, so two sub-populations with comparable abundances might be expected \citep{Kushnir2016,Hotokezaka2017,Zaldarriaga2017}. 
It has been pointed out that with different wind and tide models, less extreme distributions may result, with the possibility of having intermediate spins even after tidal locking \cite{Qin2018, Bavera2020}. We will first probe the more extreme model which is easier to constrain, keeping in mind that the bounds we obtain apply to the fraction of black holes with maximal spin, and later explore the consequences of a milder spin distribution.

Following Ref.~\cite{Roulet2019}, we implement a model of this scenario as follows. We consider that component black holes have $\chi_z \sim \mathcal N_{\rm t}(0, \sigma_\chi)$ in the absence of tidal effects, and a fraction $\zeta$ of the secondary (lighter) BHs have $\chi_z = 1$ due to a tidally locked progenitor. 
This distribution is very different from both the injection distribution and the parameter estimation prior in the space of component spins. In order to have well-behaved reweighting of samples (see \S\ref{ssec:distribution_choice}), in practice we implement it under the approximation that the strain waveform depends on the spins only through $\chieff$. We therefore use
\begin{equation}
    \begin{split}
        &f_\chieff(\chieff \mid q, \zeta, \sigma_\chi)\\
        &= \iint \rmd \chi_{1z} \rmd \chi_{2z}
            f(\chi_{1z}, \chi_{2z} \mid \zeta, \sigma_\chi) 
            \delta\Big(\chieff - \frac{
            \chi_{1z} + q \chi_{2z}}{1+q}\Big) \\
        &= (1 - \zeta)\, \mathcal N_{\rm t}(\chieff \mid \mu_0, \sigma_0) + \zeta\, \mathcal N_{\rm t}(\chieff \mid \mu_1, \sigma_1),
    \end{split}
\end{equation}
where the subscript 0 represents the case in which no tidal locking occurred and 1 the case where the progenitor of the secondary was locked:
\begin{equation}
    \begin{aligned}[c]
        \mu_0 &= 0 &
        \sigma_0 &= \frac{\sqrt{1+q^2}}{1+q} \sigma_\chi\\
        \mu_1 &= \frac q{1+q} &
        \sigma_1 &= \frac{\sigma_\chi}{1+q}.
    \end{aligned}
\end{equation}

\begin{figure}
    \centering
    \includegraphics[width=\linewidth]{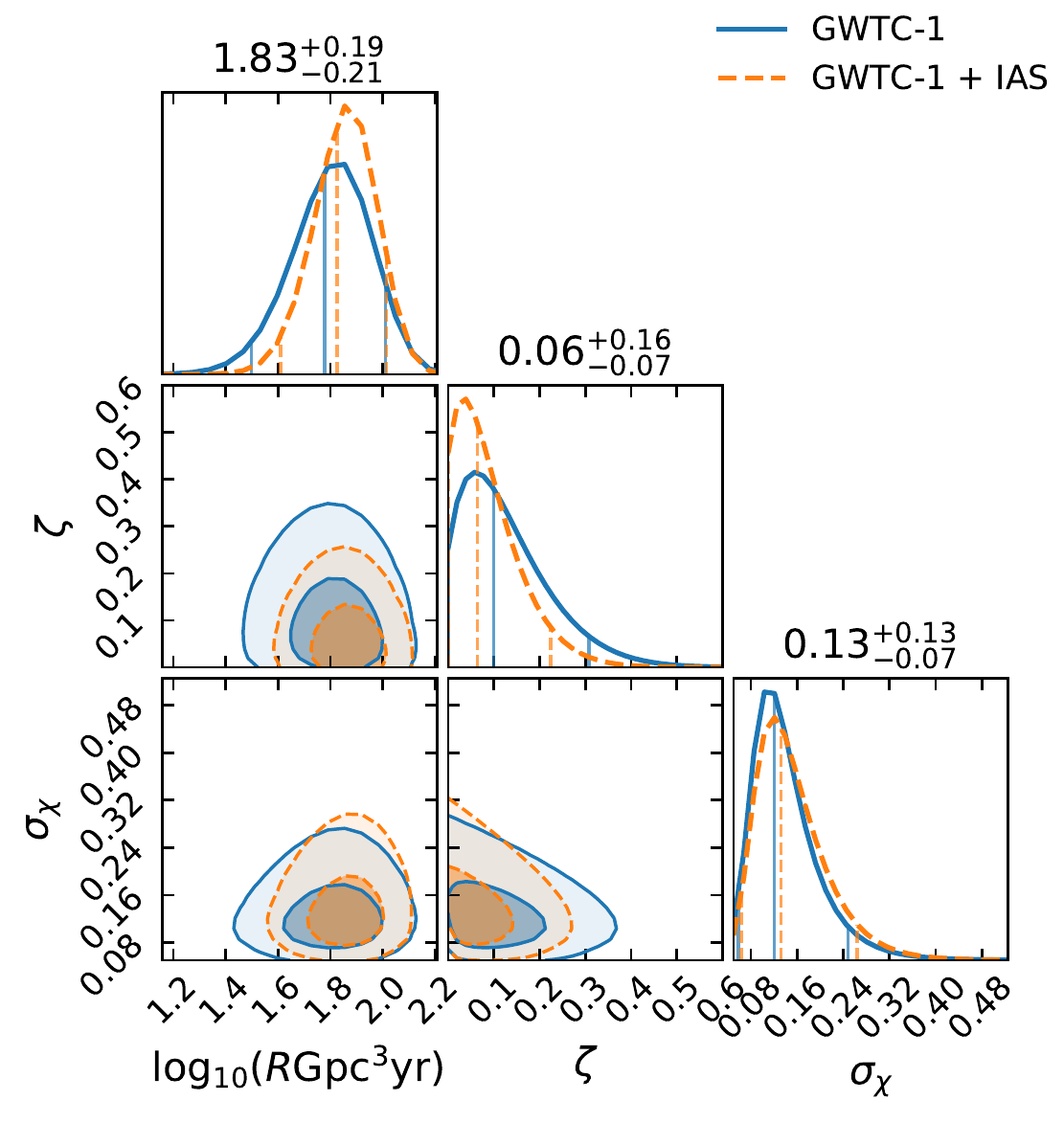}
    \includegraphics[width=.97\linewidth]{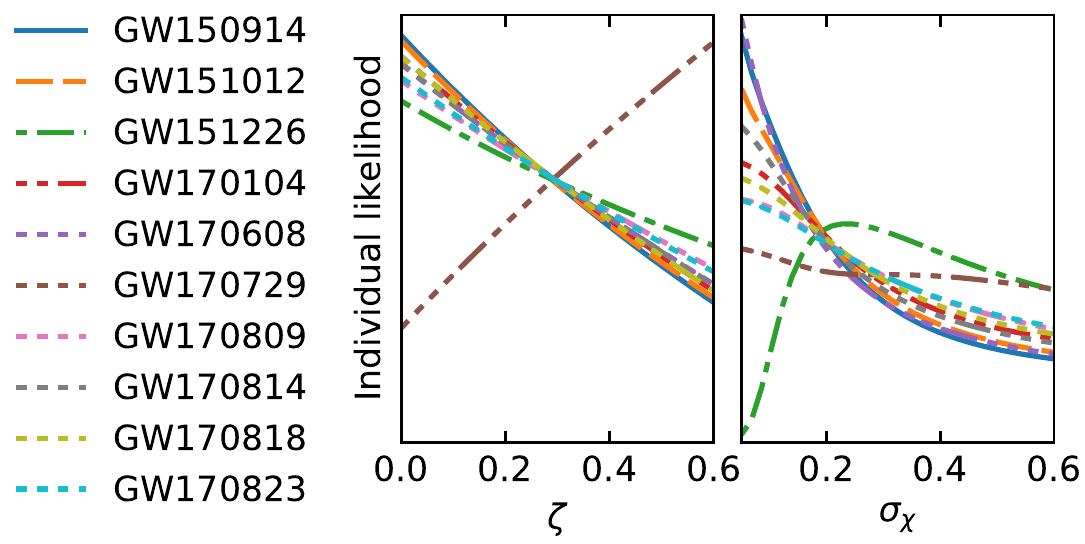}
    \caption{Model \ref{model:locked}: a fraction $\zeta$ of the secondary black holes have tidally locked progenitors ($\chi_{2z} = 1$) and the remaining $(1-\zeta)$ fraction follow the Gaussian distribution $\chi^z_{1, 2} \sim \mathcal N_{\rm t}(0, \sigma_\chi)$.\textit{ Top panel:} Population parameter constraints.\textit{ Bottom panels:} Individual marginalized likelihood for each of the events in the GWTC-1 catalog.}
    \label{fig:secondary_locked_mixture}
\end{figure}

\begin{figure}
    \centering
    \includegraphics[width=\linewidth]{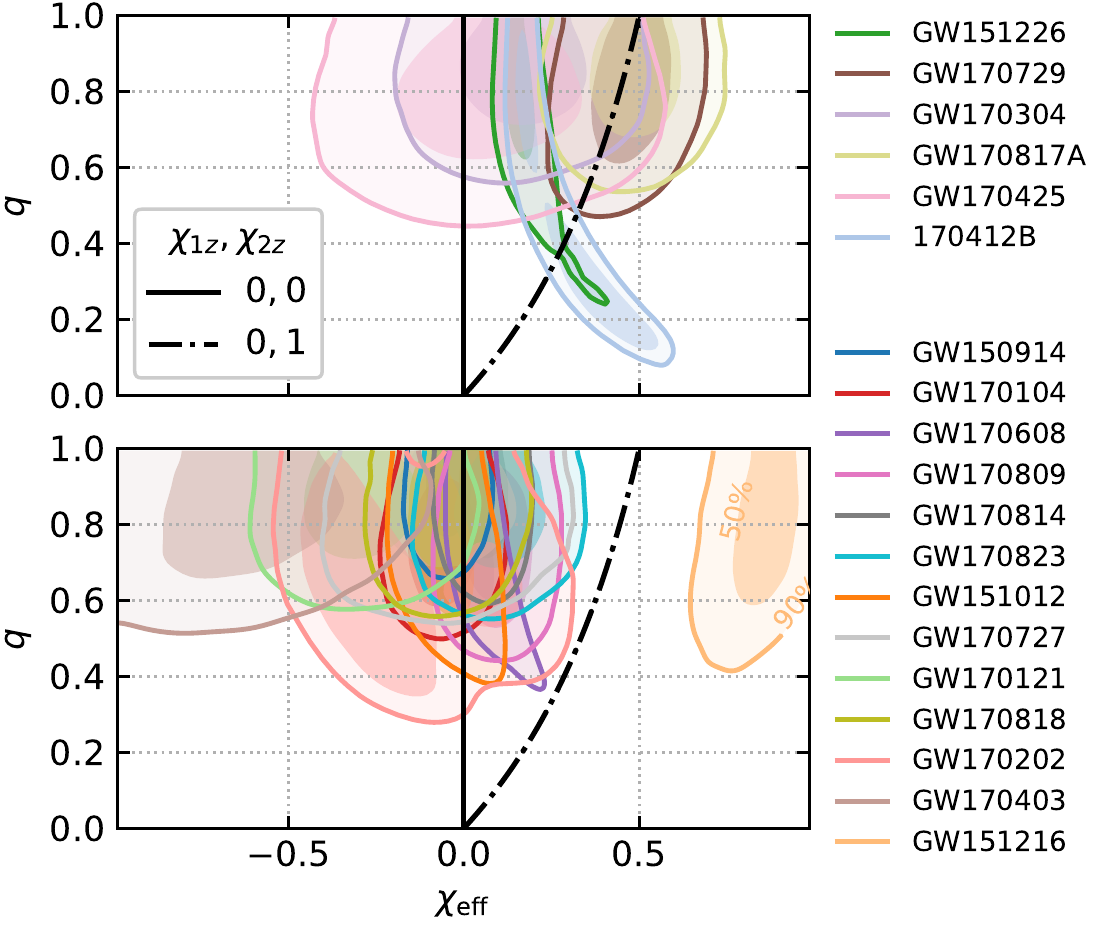}
    \caption{Posterior on the $q$--$\chieff$ plane for the events considered in this work that are consistent (top) or inconsistent (bottom) with having a non-spinning primary and a maximally spinning, aligned secondary at the 90\% confidence level.}
    \label{fig:q_chieff}
\end{figure}

We show our constraints in the top panel of Fig.~\ref{fig:secondary_locked_mixture}. As in Ref.~\cite{Roulet2019}, we find that the fraction of locked systems $\zeta$ is consistent with 0. We bind it to $\zeta < 0.2$ at 90\% confidence. Interestingly, the extreme version of this model $\sigma_\chi \approx 0$, where black holes are born with nearly zero spins except for tidal effects \cite{Fuller2019}, is in mild tension.
Even when restricted to the GWTC-1 catalog, this conclusion is in disagreement with \cite{Roulet2019}. The immediate cause of the difference is that in this work we found shorter tails in the mass-ratio distribution of some events, and these tails can affect how consistent an event is with the maximally-spinning-secondary hypothesis.\footnote{Parameter estimation samples are available at \url{github.com/jroulet/O2_samples}.}
The main differences between the parameter estimation methods of the two analyses, which could explain the disagreement, are that in this work we apply an exact treatment of the detector locations and orientations instead of the approximations made in \cite{Roulet2019} and we have a more careful estimation of the noise PSD \cite{psddrift_paper, Huang2020}, thus the new results are preferred. In Fig.~\ref{fig:q_chieff} we show the events' posteriors in the $q$--$\chieff$ plane, and the curve $\chieff = q/(1+q)$ corresponding to $\chi_{1z}=0$, $\chi_{2z}=1$ as representative of the tidally-locked progenitor scenario. In particular GW151226, whose effective spin is positive and well-measured, is only marginally consistent with having a $\chi_{2z}$ as high as 1 and prefers lower values.
In the bottom panels of Fig.~\ref{fig:secondary_locked_mixture} we show this situation in further detail. For the events in the GWTC-1 catalog, we plot the individual marginalized likelihoods (i.e. the terms inside the product in Eq.~\eqref{eq:pastro=1}) for $\zeta$ and $\sigma_\chi$. We see that the low-$\sigma_\chi$ solution is disfavored by GW151226. Additionally, GW170729 is more consistent with having $\chi_{2z}=1$ than with coming from a low spin distribution, thus it pushes the locked fraction upwards.

To keep the number of parameters small, in Model~\ref{model:locked} we have implicitly assumed that the second-formed black hole, whose progenitor can be subject to strong tides, is the lightest. While lighter stars have longer lifetimes in isolation, for binary evolution the mass ratio may be reversed by mass transfer episodes \cite{Gerosa2013, Steinle2020}. Moreover, depending on the detailed ordering of mass transfer and core collapses both stars may be subject to strong tides \cite{Steinle2020}. We explored the impact of this simplification using a model where either BH has small or high-aligned spin independently with some probability, when we do this we obtain similar constraints on the total fraction of tidally-locked progenitors and natal spin dispersion as with Model~\ref{model:locked}.

\submodel{partially_locked}
\subsubsection{Model~\ref{model:partially_locked}: tidally-torqued progenitors with moderately spinning remnants}
\begin{figure}
    \centering
    \includegraphics[width=\linewidth]{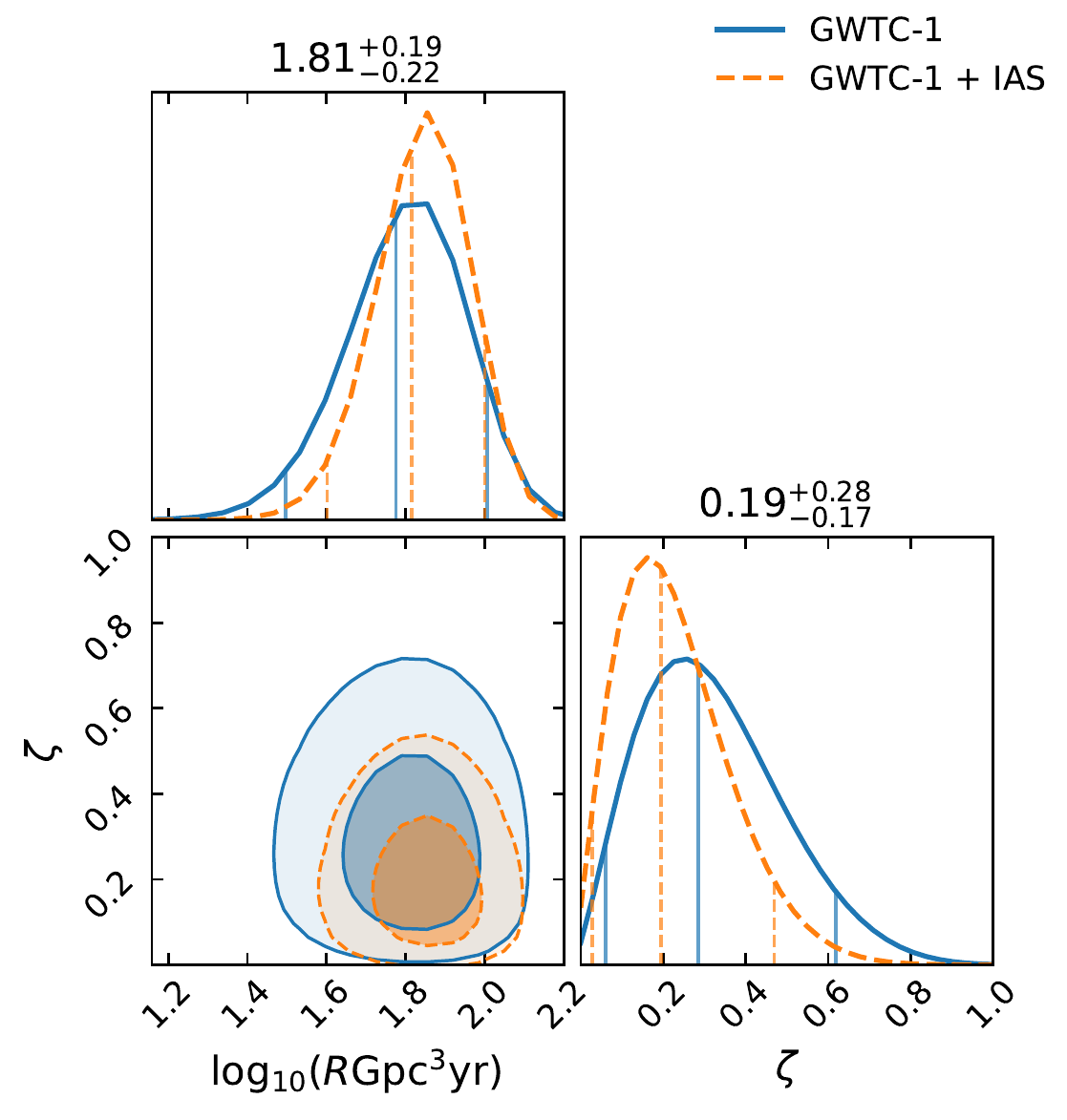}
    \caption{Model~\ref{model:partially_locked}, a modified version of Model~\ref{model:locked} where the remnant of the tidally-locked progenitor has $\chi_{2z}$ distributed uniformly in $[0, 1]$ for a fraction $\zeta$ of the mergers, and otherwise black holes have very low spins.}
    \label{fig:partially_locked}
\end{figure}

Is the tension with very low natal BH spins robust to a less extreme model for the spin of remnant of a tidally torqued star? Motivated by \cite{Qin2018, Bavera2020} we now study a version of the model where these black holes have spin aligned with the orbit, but with a broad distribution, which we take to be $P(\chi_{2z}) = \rm U(0, 1)$ for simplicity. We assume that otherwise all black holes have very low spin \cite{Fuller2019}. We parameterize the $\chieff$ distribution as
\begin{equation}
   \begin{split}
    	f_\chieff(\chieff \mid q; \zeta) &= \zeta\, {\rm U}(\chieff \mid 0, q/(1+q)) \\
    	&\quad + (1-\zeta) \,\mathcal N_{\rm t}(\chieff \mid 0, 0.05).
    \end{split} 
\end{equation}
For the subpopulation where tides did not play an important role, we have set a nonzero width in $\chieff$ comparable to measurement uncertainties, and made the approximation that the likelihood depends on the spins through $\chieff$ only, in order to avoid pathological reweighting.
We show the constraints on Fig.~\ref{fig:partially_locked}. Interestingly, under these assumptions we find that a fraction $\zeta = 0.22^{+0.29}_{-0.19}$ of the BBHs needs to have been subject to strong tidal effects in order to explain the events that are measured to have positive $\chieff$.
We have verified that if we allow for a further spread $\sigma_\chi$ in the natal BH spin distribution, this spread is well consistent with 0, unlike for Model \ref{model:locked}.

In conclusion, the data is consistent with either a natal distribution of spins with nonzero dispersion $\sigma_{\chi} \sim 0.1$, or a model where the remnant may have a moderate spin even after tidal torquing. In the first case, the result holds even if we allow a fraction of tidally-locked progenitors (the inferred fraction is consistent with zero); in the latter, which better aligns with the predictions of Refs.~\cite{Qin2018, Fuller2019, Bavera2020}, a fraction $\zeta \sim 0.2$ of tidally torqued events is favored. Later, in \S \ref{ssec:discussion}, we will show results for the relative evidence for these families of models and find some preference for Model~\ref{model:partially_locked} over \ref{model:gaussian_chieff} and \ref{model:locked}. Again, we find that including the events from the IAS catalog improves the constraints on these population models.

\model{m1_powerlaw}
\subsection{Model \ref{model:m1_powerlaw}: truncated power-law in primary mass}
Next we consider a model where the source-frame mass of the primary black hole follows a truncated power-law distribution
\begin{align} \label{eq:f_m1}
    f_\mones(\mones \mid \alpha, m_{\rm min}, m_{\rm max}) &\propto m_{\rm 1s}^{-\alpha};
    ~  m_{\rm min} < \mones < m_{\rm max} \\
    f_q(q \mid \mones, m_{\rm min}) &= {\rm U}(m_{\rm min} / \mones, 1) \label{eq:fq_m1}.
\end{align}
Eq.~\eqref{eq:fq_m1} enforces that the secondary mass also satisfies the lower cutoff.
This model was first studied by Ref.~\cite{Fishbach2017} (see also \cite{Roulet2019, LVC_pop_O2}) and is motivated by the prediction of a gap in the stellar black hole mass function due to the pair instability supernova and pulsational pair instability supernova processes \cite{Fowler1964, Barkat1967, Bond1984, Heger2003}.

\begin{figure}
    \centering
    \includegraphics[width=\linewidth]{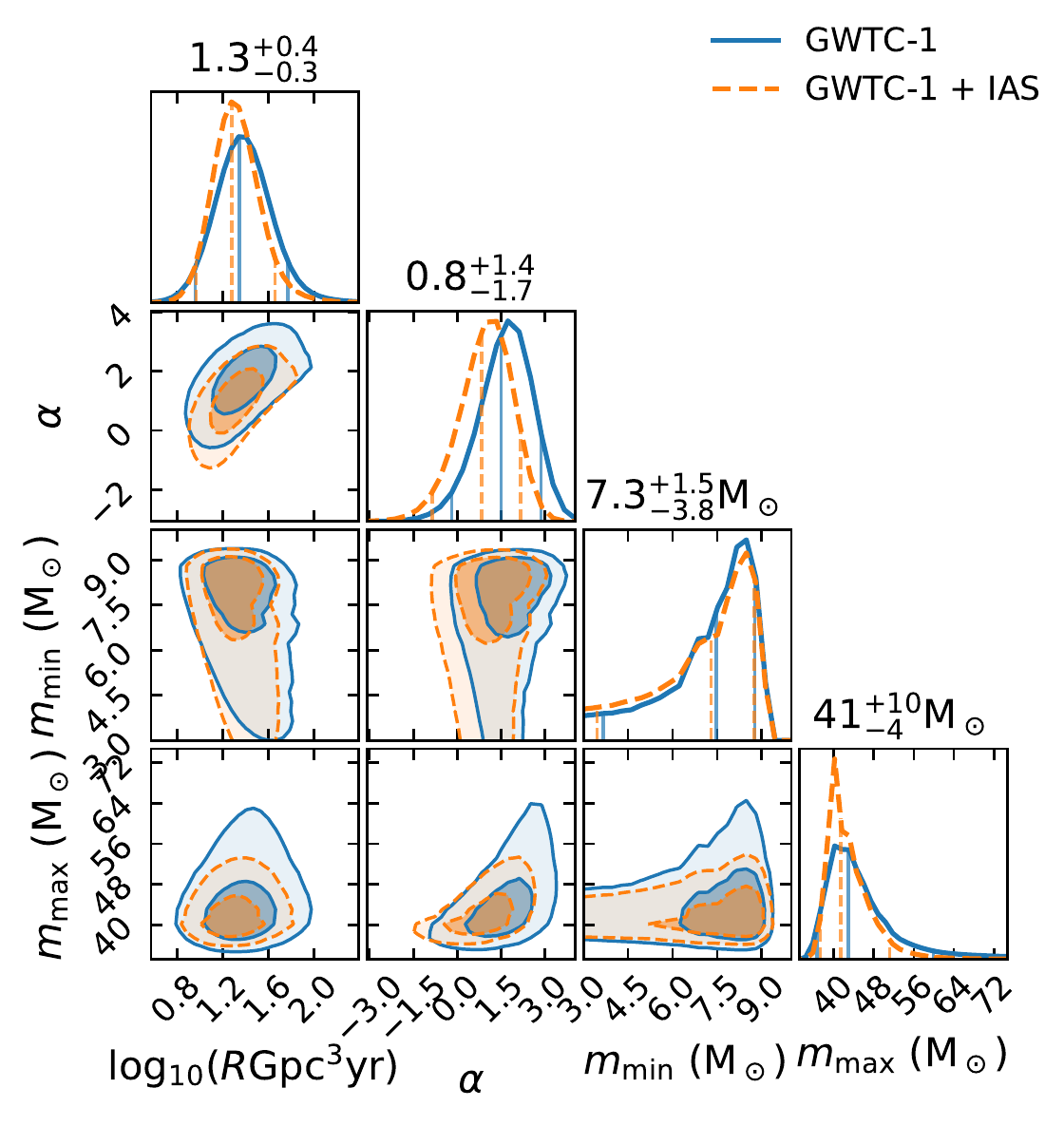}
    \caption{Model \ref{model:m1_powerlaw}: a power-law distribution for the primary black hole mass truncated at the minimum and maximum masses.}
    \label{fig:m1_powerlaw_bothcutoff}
\end{figure}

We show our constraints in Fig.~\ref{fig:m1_powerlaw_bothcutoff}. These might be approximately compared to the models explored in \cite[figure 2]{LVC_pop_O2}. When restricting to the events in GWTC-1 we obtain similar results to Ref.~\cite{LVC_pop_O2}, except we recover a merger rate that is lower by $\sim 0.4$ decades, which is smaller than the current statistical uncertainty.
Interestingly, we obtain a tighter constraint $m_{\rm max} < \SI{51}{M_\odot}$ at 90\% confidence by including the events from the IAS catalog. This is accompanied by a slight shift in the allowed power-law index towards shallower slopes. A putative lower cutoff $m_{\rm min}$ in the mass function is harder to detect since low-mass mergers are intrinsically fainter. Thus, similar to \citet{LVC_pop_O2}, we can only put an upper bound on $m_{\rm min}$ given by the lightest confident merger considered, GW170608.

We comment that the O3 event GW190521 has a primary black hole mass $85^{+21}_{-14}\,\rm M_\odot$ \cite{GW190521}, in tension with the constraints in Fig.~\ref{fig:m1_powerlaw_bothcutoff}. This suggests that the truncated power-law model may not be a good description of the tails of the distribution once a larger number of events is included, and that parametrizations with more complexity might be needed in future analyses.

\model{q_powerlaw}
\subsection{Model \ref{model:q_powerlaw}: power-law in the mass ratio}
\begin{figure}
    \centering
    \includegraphics[width=\linewidth]{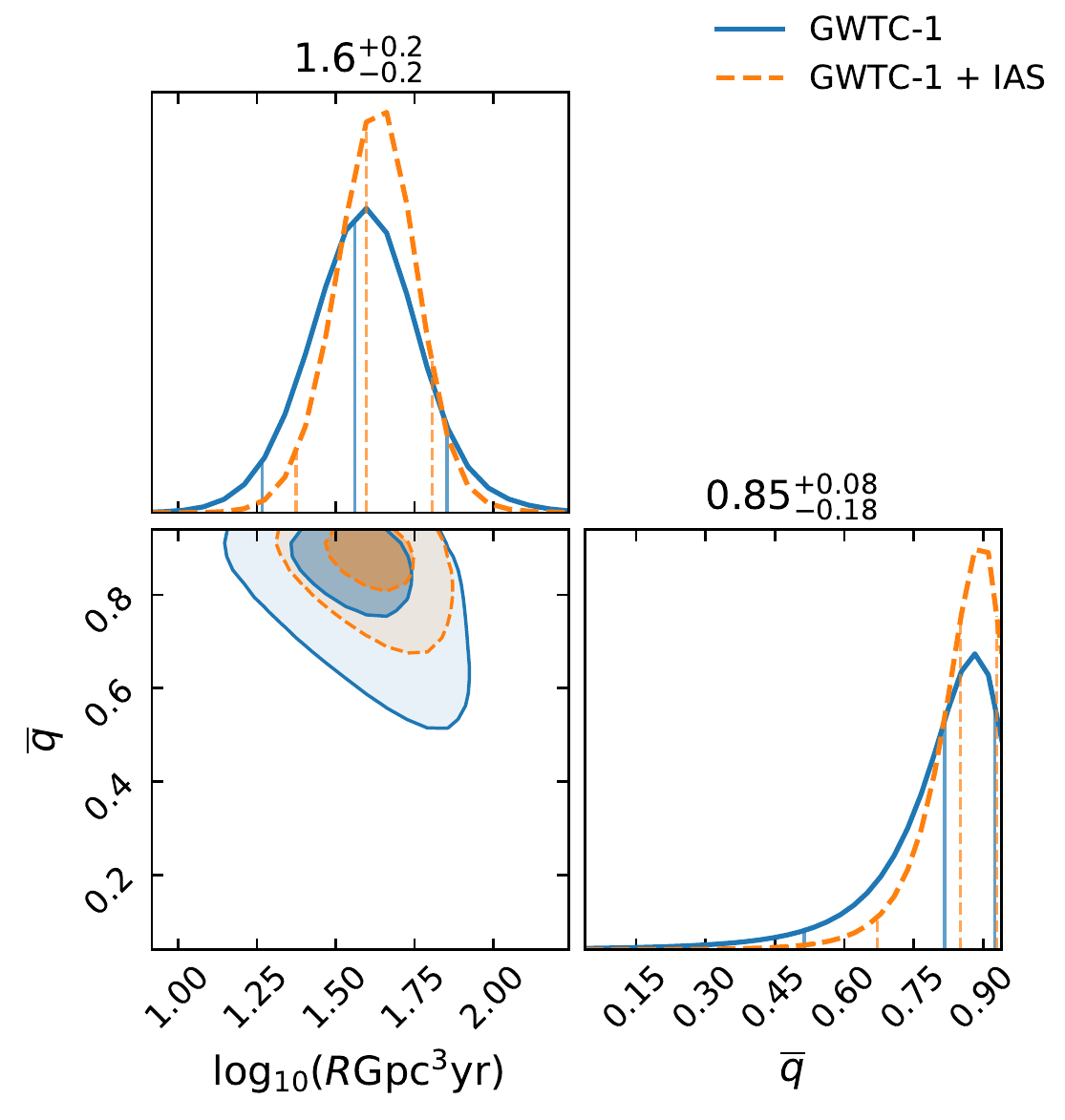}
    \caption{Model \ref{model:q_powerlaw}: a power-law in the mass ratio with mean $\overline q$, Eq.~\eqref{eq:q_powerlaw}.}
    \label{fig:q_powerlaw}
\end{figure}

Now we study a model where the mass ratio follows a power law distribution \cite{Roulet2019, Fishbach2020}
\begin{equation} \label{eq:q_powerlaw}
    f_q(q \mid \overline q) \propto q^\beta
\end{equation}
with $\beta = (2\, \overline q - 1) / (1 - \overline q)$ so that the distribution has a mean $\overline q$. We show our constraints in Fig.~\ref{fig:q_powerlaw}. In line with previous results \cite{Roulet2019, LVC_pop_O2, Fishbach2020}, we find that distributions leaned towards equal mass ratios are favored. Including the events from the IAS catalog enables a more precise measurement of both the rate and mean mass-ratio. We find a mild quantitative difference in the $\overline q$ distribution with \cite[figure 9]{Roulet2019}, which we verified is due to the difference in the underlying mass distribution models (truncated power-law in primary mass per Eq.~\eqref{eq:fm1} vs. flat in chirp mass);  the one used here is strongly favored in terms of model selection by a difference in log-likelihood $\Delta \max_\lambda \ln P(\{d_i\}, \ntrig \mid \lambda) = 24$.

\newcounter{model}

\model{z_powerlaw}
\subsection{Model \ref{model:z_powerlaw}: power-law in the redshift}
\label{ssec:redshift}
\begin{figure}
    \centering
    \includegraphics[width=\linewidth]{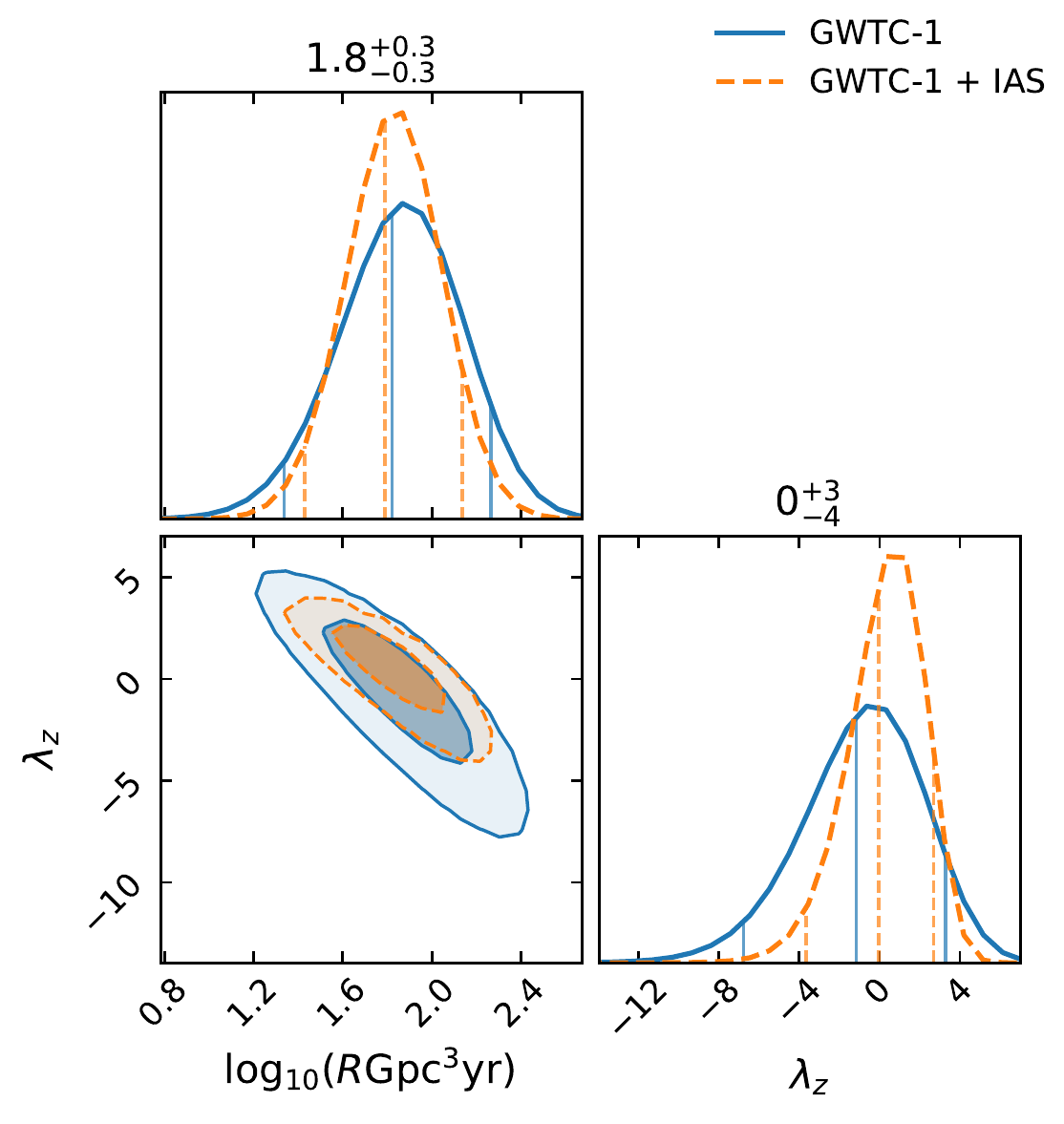}
    \caption{Model \ref{model:z_powerlaw}: a power-law in the redshift evolution of the merger rate with exponent $\lambda_z$, Eq.~\eqref{eq:z_powerlaw}. $R$ is the local rate.}
    \label{fig:z_powerlaw}
\end{figure}

Finally, we consider a model where the comoving merger rate follows a power-law in the redshift with index $\lambda_z$ \cite{Fishbach2018}
\begin{equation} \label{eq:z_powerlaw}
    f_{D_L}(D_L \mid \lambda_z) = \hat f_{D_L}(D_L)(1+z)^{\lambda_z},
\end{equation}
so that $\lambda_z = 0$ corresponds to a constant merger rate per unit comoving volume (see Eq.~\eqref{eq:fD}). We show the constraints on $\lambda_z$ in Fig.~\ref{fig:z_powerlaw}: it is poorly constrained and is consistent with a constant merger rate per unit comoving volume. Adding the new events from the IAS catalog improves the constraints both on the rate and on $\lambda_z$.

\subsection{Merger rate} \label{ssec:rate}

As seen from Figs.~\ref{fig:gaussian_chieff}--\ref{fig:z_powerlaw}, the measurements of the local merger rate $R$ using each of the models we studied are consistent within uncertainties, however, they have broad distributions and may have large correlations with some of the population parameters. This happens because the rate parameter is measured through the combination $N_a(\lambda) = R \cdot \overline{VT}(\lambda')$, and the population-averaged sensitive-volume $\overline{VT}(\lambda')$ can exhibit a large dependence on the population shape. In other words, most of the information comes from the region in parameter space where most events lie. This region depends on the interplay between the astrophysical population and the detector selection function; the rate within this region should be relatively well constrained compared to the overall rate.

The most extreme example is Model~\ref{model:z_powerlaw} (power-law in the redshift, Fig.~\ref{fig:z_powerlaw}), where the rate exhibits a strong correlation with the exponent $\lambda_z$. In this case, the merger rate at some intermediate redshift $z\sim0.2$ is much better constrained than the local rate $R$, owing to the larger phase space that the detectors are sensitive to. To various extents, a similar effect occurs with some shape parameters in the other models studied in this section. In Model~\ref{model:m1_powerlaw} (power-law in the primary mass, Fig.~\ref{fig:m1_powerlaw_bothcutoff}), for steep power-laws the rate can be dominated by the low-mass end of the distribution, which is poorly constrained because the sensitive volume to these signals is smaller. 

\begin{figure}
    \centering
    \includegraphics[width=\linewidth]{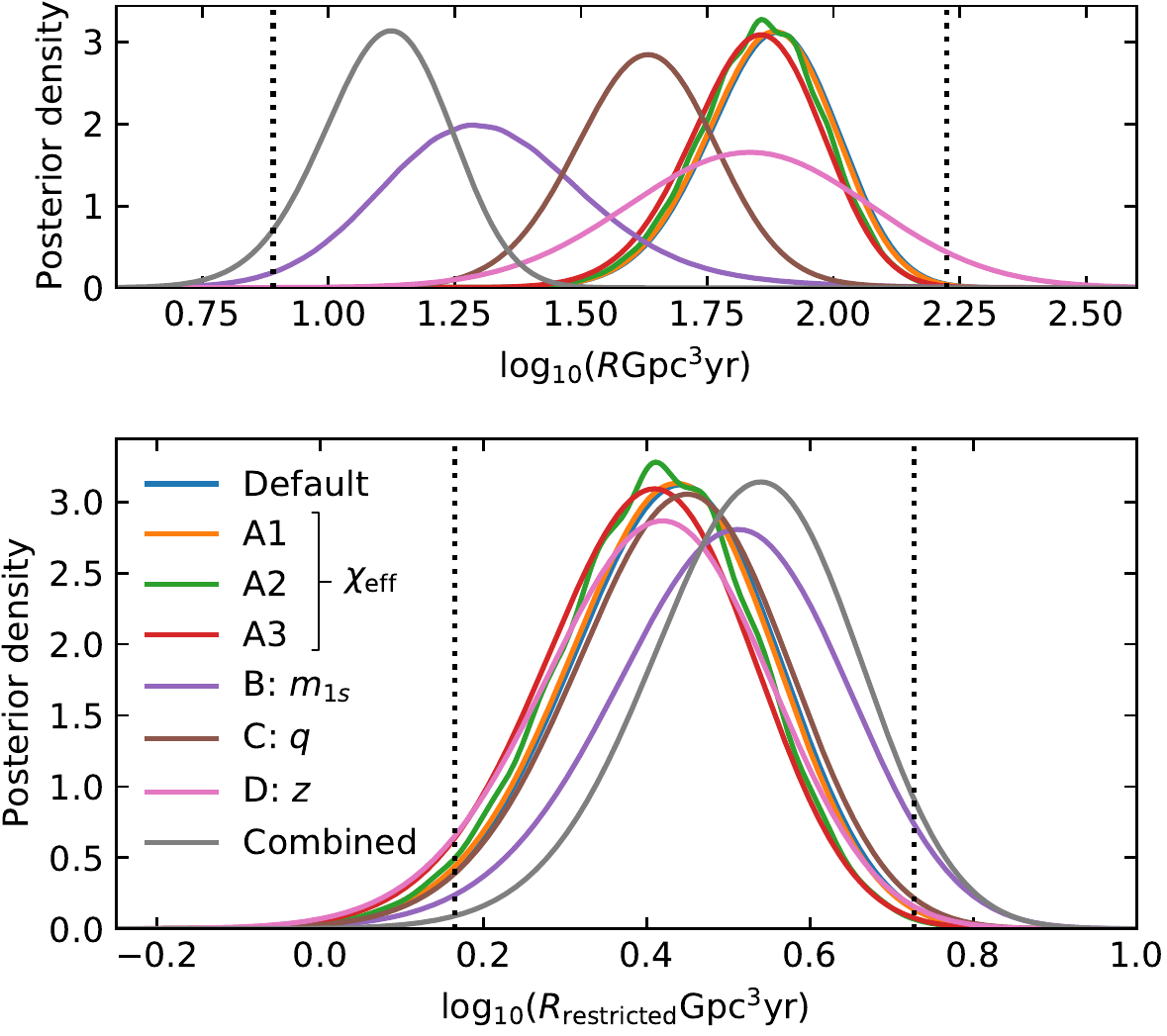}
    \caption{Posterior for the local rate of mergers (top) or the comoving rate restricted to $\SI{20}{M_\odot}<\mones<\SI{30}{M_\odot}$, $q>0.5$ and evaluated at redshift $z=0.2$ (bottom), under the various models we studied (\S\ref{ssec:gaussian}--\ref{ssec:redshift}, annotated with the variable each of them is intended to explore). We add a model that combines the maximum likelihood solutions of models \ref{model:gaussian_chieff}, \ref{model:m1_powerlaw} and \ref{model:q_powerlaw}. The restricted rate is better measured, and in better agreement across models, than the total local rate $R$. Dotted lines indicate the minimum over models of the 5\% quantile, and the maximum of the 95\% quantiles.}
    \label{fig:rate}
\end{figure}

A simple prescription to get a robust constraint that can inform theoretical models is to measure the rate of mergers within the part of parameter space where most events were observed. In Fig.~\ref{fig:rate} we plot the posterior distribution for the restricted rate of mergers $R_{\rm restricted}$, which we define as the merger rate of events with $\SI{20}{M_\odot}<\mones<\SI{30}{M_\odot}$, $q>0.5$, evaluated at redshift $z=0.2$. We show this quantity for the default model (Eq.~\eqref{eq:fhat}), Models \ref{model:gaussian_chieff}--\ref{model:z_powerlaw} (\S\ref{ssec:gaussian}--\ref{ssec:redshift}), and a `Combined' model that combines the maximum likelihood solutions of models \ref{model:gaussian_chieff}, \ref{model:m1_powerlaw} and \ref{model:q_powerlaw}, namely the product of $f_\chieff$ from Eq.~\eqref{eq:f_chieff} with $\overline\chieff=0, \sigma_\chieff=0.1$; $f_{\mones}$ from Eq.~\eqref{eq:f_m1} with $\alpha=1, m_{\rm min}=\SI{8.5}{M_\odot}, m_{\rm max}=\SI{40}{M_\odot}$; $f_q$ from Eq.~\eqref{eq:q_powerlaw} with $\overline q = 0.88$; and $\hat f_{D_L}$ from Eq.~\eqref{eq:fD}. Note that the primary mass is distributed with a truncated power-law of index $-2.35$ in all except the \ref{model:m1_powerlaw} and Combined models, and for the Combined model it is flat-in-log, which makes Fig.~\ref{fig:rate} largely comparable to \cite[fig.~12]{GWTC-1}; this may also be the main driver of the residual discrepancies between restricted rates across models in Fig.~\ref{fig:rate}.

We find that the restricted rate is much better measured than the total local merger rate and that all models largely agree on its value. Taking the union of the symmetric 90\% quantiles, we obtain that the rate lies in the range 1.5--\SI{5.3}{Gpc^{-3}yr^{-1}}, which amounts to a factor 3.6 uncertainty in the restricted rate as opposed to 21 in the absolute rate, within the models we explored. Thus, the restricted rate is well suited to put population models to a more stringent test.

\subsection{Comparison between population models}\label{ssec:discussion}

We conclude this section by comparing the performances of the models we studied. In Table~\ref{tab:model_scores} we provide the maximum likelihood and the Bayesian evidence achieved by models \ref{model:gaussian_chieff}--\ref{model:z_powerlaw} and Combined discussed above, relative to the default model Eq.~\eqref{eq:fhat}.
The maximum likelihood has the advantage of being independent of the arbitrary prior choices for the population parameters (in particular, their ranges), on the other hand it does not penalize models with more degrees of freedom. The ordering and qualitative conclusions are found to be largely similar using either metric.
Our method involves multiple Monte Carlo integrations (numerator and denominator of Eq.~\eqref{eq:wi_est}, and Eq.~\eqref{eq:g_est}), which can introduce stochastic errors. We estimate these with the bootstrap method, by computing the scores 100 times using samples taken with replacement from the original sets.
\begin{table}
\begin{tabular}{lrr}
\tabline
Model & $\Delta \max \ln L$ &       $\Delta \ln Z$ \\
\tabline
Default     &                        $0$ &                      $0$ \\
\multirow{3}*{\hspace{-4pt}$\begin{array}{l}\rm A1\\\rm A2\\\rm A3\end{array}\Bigg\}\chieff$} & $14.6_{-1.1}^{+1.3}$ &     $10.3_{-0.8}^{+1.1}$ \\
          &       $14.9_{-0.9}^{+1.3}$ &     $12.0_{-0.8}^{+1.1}$ \\
          &       $14.6_{-1.2}^{+1.5}$ &     $13.4_{-1.1}^{+1.4}$ \\
B: $m_{1s}$ &        $3.8_{-0.5}^{+0.6}$ &     $-2.1_{-0.3}^{+0.5}$ \\
C: $q$      &        $4.6_{-0.6}^{+0.6}$ &      $2.8_{-0.4}^{+0.5}$ \\
D: $z$      &     $0.07_{-0.06}^{+0.19}$ &  $-2.84_{-0.12}^{+0.29}$ \\
Combined    &             $22_{-3}^{+3}$ &           $22_{-2}^{+2}$ \\
\tabline
\end{tabular}
\caption{Scores for the population models we study (\S\ref{ssec:spin}--\ref{ssec:redshift}, annotated with the variable each of them is intended to explore) relative to the Default model, Eq.~\eqref{eq:fhat}. We add a model that combines the maximum likelihood solutions of models \ref{model:gaussian_chieff}, \ref{model:m1_powerlaw} and \ref{model:q_powerlaw}. We report the maximum likelihood and the Bayesian evidence as complementary indicating scores. Error-bars indicate 90\% confidence levels on the uncertainties from the Monte Carlo method employed.}
\label{tab:model_scores}
\end{table}

We find that among Models \ref{model:gaussian_chieff}--\ref{model:z_powerlaw}, those that perform best are \ref{model:gaussian_chieff} (Gaussian in $\chieff$), \ref{model:locked} and \ref{model:partially_locked} (tidally locked progenitors, with $\chi_{2z}=1$ or $\chi_{2z}\sim\rm U(0, 1)$ respectively), which vary the spin distribution away from the flat-in-$\chieff$ one the default model uses. This is a clear indication that the average effective spins of the population are lower. The three models achieve similar likelihoods, Model \ref{model:partially_locked} has a somewhat higher Bayesian evidence, which can be related to the fact that it has one less parameter. Note that these models cannot accommodate the high effective spin of GW151216, which requires both spins to be high and aligned; its $\overline \pastro$ is suppressed as a result (Table~\ref{tab:pastro}). To a lesser extent, a similar effect holds for GW170817A and GW170403.

The Combined model, which is defined from the best likelihood solutions of Models \ref{model:gaussian_chieff}, \ref{model:m1_powerlaw} and \ref{model:q_powerlaw}, indeed outperforms its individual components and, as expected, the likelihood ratio to the default model is approximately the product of the component likelihood ratios to the default model. This vindicates our approach of individually varying the components of the population model. 

\begin{figure}
    \centering
    \includegraphics[width=\linewidth]{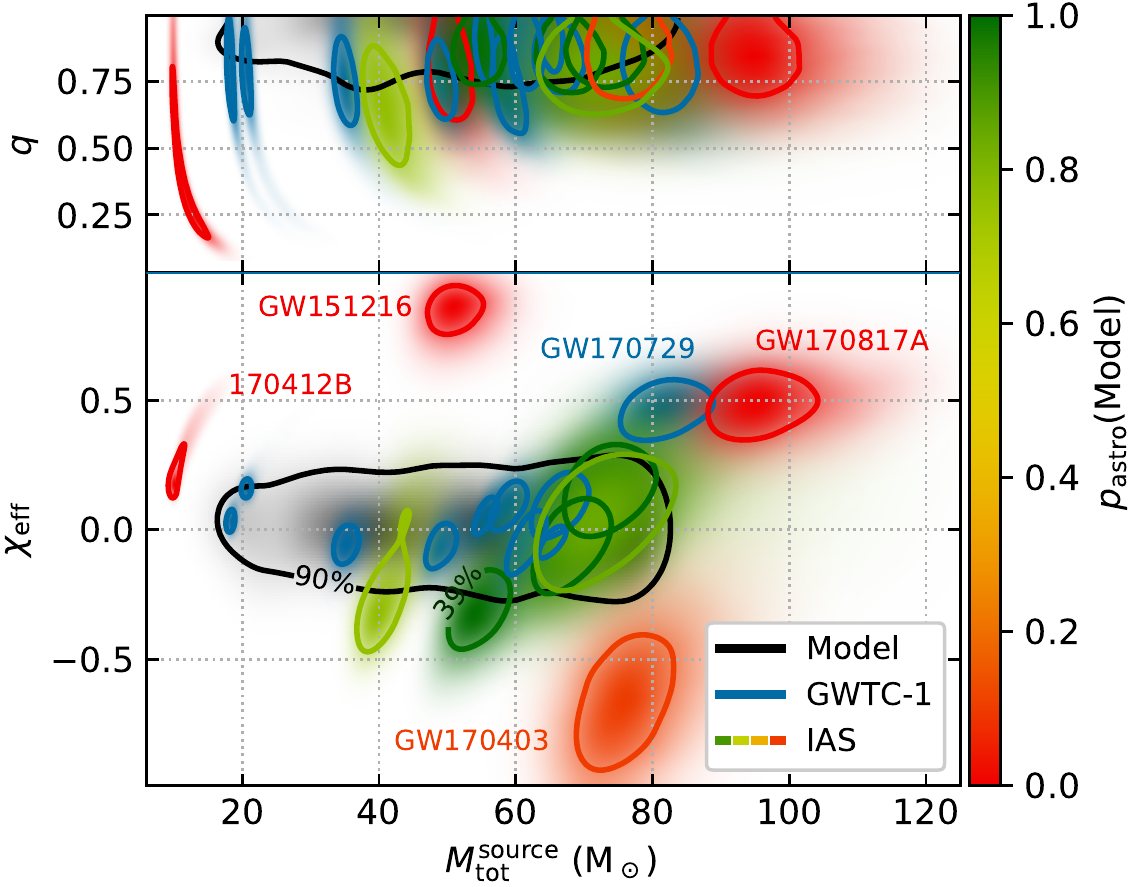}
    \caption{Events considered in this work as a function of total source-frame mass, mass-ratio and effective spin. Underlaid in black is the prediction of a model that combines the maximum likelihood solutions for the effective spin (Model~\ref{model:gaussian_chieff}), primary mass (Model~\ref{model:m1_powerlaw}) and mass-ratio (Model~\ref{model:q_powerlaw}) distributions, including selection effects and without measurement uncertainty. The events' posteriors are color-coded according to their values of $\pastro$ under the same model (these are different from the values reported by the search pipeline). Note however that these posteriors are computed using a prior that differs from the model. Model outliers are labeled.}
    \label{fig:events}
\end{figure}

In Fig.~\ref{fig:events} we plot the BBH events from O1 and O2 considered in this work, showing posteriors in the space of total source-frame mass versus mass-ratio and effective spin.
$1\sigma$ contours (enclosing $1-e^{-1/2} \approx 0.39$ of the distribution) are drawn in blue for events in the GWTC-1 catalog or color-coded by $\pastro$ for events in the IAS catalog. 
The broad parameter estimation prior $\pi(\theta)$, defined in \S\ref{ssec:distribution_choice}, is used for event contours in order to make the resulting posteriors trace more closely the single-event likelihoods from Eq.~\eqref{eq:Gaussian_like}.
Shown in black is the expected distribution of detectable sources under the Combined model, with a 90\% contour. It is obtained by reweighting found injections with this model.
The $\pastro$ values shown in the color scale of Fig.~\ref{fig:events} correspond to this model. Note that this is just one example out of the set of models consistent with the data, and others may exhibit somewhat different behavior.

This simple model appears to explain the observations reasonably well, albeit with some outliers. The confident event GW170729 \cite{GWTC-1, BBH_O2} is a mild outlier within $2\sigma$ of the 90\% contour, note however that under this population model its posterior would shift towards lower mass and $\chieff$ \cite{Fishbach2020b}. The other outliers (GW170817A, GW170403, GW151216 and 170412B) have less detection significance, so their $\pastro(\lambda)$ can be suppressed without too large a penalty to the model likelihood.
As our knowledge of the astrophysical distribution improves, the $\pastro$ of these events might be revised.
A potential limitation of the models studied in this work is the assumption that the distribution is factorizable in the variables $\mones, q, \chieff, D_L$ (Models \ref{model:locked} and \ref{model:partially_locked} do have a correlation between $q$ and $\chieff$, and Model \ref{model:m1_powerlaw} between $\mones$ and $q$). Indeed, Fig.~\ref{fig:events} might hint that models in which the $\chieff$ mean or variance depends on the mass \cite{Safarzadeh2020} or distance \cite{Hotokezaka2017} could perhaps better accommodate some of these outliers. Looking forward, as the catalog of events grows there might be a need for increasingly complex models.
 
We comment that 170412B is consistent with having a secondary mass similar to that of the O3 event GW190814 \cite{GW190814}, although its $\chieff$ would be higher especially if such a small mass-ratio is imposed.

\section{Conclusions}
\label{sec:conclusions}

In this work, we presented a general framework to include marginal GW events when characterizing the astrophysical population of BBH mergers. 
Similar to other proposed methods, ours requires that we characterize the sensitivity of the search pipeline. 
We achieved this using software injections, which we generated using a fiducial distribution and reweighted as needed. 
We generated the parameters of the injected events from a reference population model weighted by an approximate detector sensitivity function; this simple prescription achieved a good balance between accepted and rejected injections. 
We implemented an injection campaign using the above strategy, and empirically measured the sensitive volume-time of our search to be $\simeq$ \SI{0.17}{Gpc^3\,yr}, for the population model in Eq.~\eqref{eq:fhat} and with a FAR threshold of one per all our O1 and O2 BBH searches combined (Appendix \ref{ssec:vt}). In its own right, quantifying the sensitivity of our pipeline solidifies the evidence in favor of the detections that we previously reported.

We demonstrated our method by using events reported in the IAS catalog, in addition to the ones originally reported by the LVC, to characterize the BBH population. 
In particular, we studied various phenomenological population models that explored the spin, mass, mass-ratio and redshift distributions. 
For models that have been previously explored, our results are broadly consistent with previous studies, with reduced uncertainties due to the extra information that the additional events contribute. We quantified the information gain from including these additional events, e.g. for the astrophysical rate, it scales with their summed $p_{\rm astro}^2$ and amounts to a $\sim 47\%$ increase. 

In models where the effective spin parameter, $\chieff$, of all mergers is drawn from a normal distribution, we do not find any statistically significant deviations from $\overline \chieff = 0$, and infer a typical spread of $\sigma_{\chi_{\rm eff}} \sim 0.1$. 
If we allow a fraction $\zeta$ of the secondary black holes to have aligned spins due to tidal effects on their progenitors, the conclusions depend on how efficient the tides can be. If the tides, when operative, are strong enough that the secondary BBH ends up with maximal spin, the fraction $\zeta$ is consistent with zero and bounded to $\zeta < 0.2$ (the spread $\sigma_{\chi_{\rm eff}}$ of the mergers without tides is similar to the previous case). 
If the tides are weak enough or the details of the collapse allow for an aligned, but not necessarily maximally-spinning secondary, the data can be explained by a fraction $\zeta \sim 0.2$ of BBHs with tidally torqued progenitors, with the rest having very low natal spins. 
Future data might be able to distinguish between these two scenarios; it is intriguing that two of the BBH mergers reported so far in the O3 run have non-zero and positive values of $\chieff$ \cite{GW190412, GW190814}.

The mass, mass-ratio, and redshift distributions are consistent with previous work: if the masses are drawn from a truncated power-law distribution, we bound the upper cutoff in the primary mass to $m_{\rm max} < \SI{51}{M_\odot}$ at 90\% confidence. The data favor a mass-ratio distribution that leans towards the equal-mass case, $\overline q > 0.67$, and a redshift distribution that is consistent with uniform in comoving volume.

We additionally argued that the merger rate is better measured if restricted to the region of parameter space where most events are found. We find that the merger rate restricted to BBHs with a primary mass between 20--\SI{30}{M_\odot}, mass ratio $q > 0.5$, and at $z \sim 0.2$, is 1.5--\SI{5.3}{Gpc^{-3} yr^{-1}} (90\% c.l.). Unlike for the total local merger rate, this constraint is model independent and a factor of $\sim 3$ tighter, and thus well-suited for testing progenitor models.

Apart from the results on the population models for the data we included, we foresee that the methodology presented here will continue to prove useful as future data releases will generically include marginal detections. We have emphasized the dependence of $\pastro$ on the astrophysical model and the search pipeline used.
An intermediate step in our method is to compute the $\pastro$ of the triggers of interest for a specified reference source population model, which we make as permissive as possible to facilitate reweighting. 
Looking forward, this can be a convenient convention for reporting values of $\pastro$, especially for marginal triggers whose final interpretation may depend on the population model.

\section*{Acknowledgements}

We thank Seth Olsen for discussion and comments on the manuscript.
We thank Daniel Wysocki for pointing us to useful references.
JR thanks the Center for Computational Astrophysics for hospitality while part of this work was carried. TV and LD acknowledge support by the John Bahcall Fellowship at the Institute for Advanced Study. This material is based upon work supported by the National Science Foundation under Grant No. 2012086.
BZ acknowledges the support of the Peter Svennilson Membership Fund and the Frank and Peggy Taplin membership fund.
MZ is supported by NSF grants PHY-1820775 the Canadian Institute for Advanced Research (CIFAR) Program on Gravity and the Extreme Universe and the Simons Foundation Modern Inflationary Cosmology initiative.

This research has made use of data, software and web tools obtained from the Gravitational Wave Open Science Center (\url{https://www.gw-openscience.org}), a service of LIGO Laboratory, the LIGO Scientific Collaboration and the Virgo Collaboration. LIGO is funded by the U.S. National Science Foundation. Virgo is funded by the French Centre National de Recherche Scientifique (CNRS), the Italian Istituto Nazionale della Fisica Nucleare (INFN) and the Dutch Nikhef, with contributions by Polish and Hungarian institutes.

\appendix

\section{Sensitivity of the search pipeline} \label{app:sensitivity}
In this appendix we report results of our injection campaign and characterize the sensitivity of our primary search for BBH mergers \cite{pipeline, BBH_O2}.

\subsection{Injection campaigns}
\label{sec:inj_campaign}

We make \num{50000} software injections in each of the O1 and O2 observing runs, at random times without regard to the duty cycle of the detectors, with source parameters distributed according to Eq.~\eqref{eq:pinj}.
We then run all stages of our search as described in Refs.~\cite{pipeline, BBH_O2}, except for the following two modifications.

First, for the injection campaigns we disable the initial stage of noise transient (glitch) rejection and inpainting \cite{pipeline, psddrift_paper}, which would otherwise greatly increase the computational cost. Instead, we keep track of the locations where glitches were identified in the original search. The pipeline does not record triggers within \SI{1}{\second} of an identified glitch for templates shorter than \SI{10}{\second}, so we treat those times as invalid for observation.

Second, we implement an improved version of the coherent score---this is the piece of the detection statistic that accounts for signal coherence across detectors \cite{Nitz2017, pipeline}, and depends on the difference in the arrival times, as well as the relative phase between the detectors. 
Our implementation of the coherent score uses the best measured values of these parameters from the data, and accounts for the requisite amounts of measurement noise in each of these values. 
However, in previous work, we had neglected the effect of the correlation between the measurements of the arrival times and phases that are input to the coherent score; we found that this ultimately caused us to assign a high FAR to the LVC event GW170818 in our coincidence search (using only Hanford and Livingston data;\footnote{Our search pipeline is so far restricted to LIGO Hanford and Livingston data, on which all our significance estimates are based. In this work we used Virgo data only for parameter estimation of events whenever these are available.} the FAR was however not biased since this effect impacted the timeslides as well). 
The new version we use in this paper accounts for the correlations in the measurements, and hence should be closer to optimality (in the sense of being closer to the likelihood ratio test). 
When we apply this improvement to the coincidence search itself (both the zero-lag triggers and those obtained using timeslides), we obtain a higher significance for GW170818 than our previous result \cite{BBH_O2} (see also \cite{fishing}). This also changes the values of the IFAR of the rest of the candidates, as we report in Table~\ref{tab:ifar}.

\begin{table}
    \centering
    \begin{tabular}{llrr}
        \tabline
             Run & Event &      GPS time & IFAR (run) \\
        \tabline
    \multirow{4}{*}{O1}&    GW150914 & 1126259462.41 &     ${}>\num{20000}$ \\
    &    GW151012 & 1128678900.43 &     ${}>\num{20000}$ \\
    &    GW151226 & 1135136350.59 &     ${}>\num{20000}$ \\
    &    GW151216 & 1134293073.16 &      $26.70$ \\
        \hline
    \multirow{13}{*}{O2}&    GW170823 & 1187529256.50 &     ${}>\num{20000}$ \\
    &    GW170809 & 1186302519.74 &     ${}>\num{20000}$ \\
    &    GW170729 & 1185389807.31 &     ${}>\num{20000}$ \\
    &    GW170814 & 1186741861.52 &     ${}>\num{20000}$ \\
    &    GW170104 & 1167559936.58 &     ${}>\num{20000}$ \\
    &    GW170727 & 1185152688.02 &     $256.41$ \\
    &    GW170121 & 1169069154.57 &     $185.19$ \\
    &    GW170304 & 1172680691.36 &      $78.74$ \\
    &    GW170818 & 1187058327.08 &      $30.40$ \\
    &    170412B\footnote{This is a new candidate we had previously missed due to an execution error that affected $\approx 0.2\%$ of O2 data, and is not to be confused with 170412 \cite{GWTC-1}. Its detector-frame chirp-mass is $\mathcal M \approx \SI{4.59}{M_\odot}$.}  & 1175991666.07 &       $6.51$ \\
    &    GW170403 & 1175295989.22 &       $6.25$ \\
    &    GW170425 & 1177134832.18 &       $5.30$ \\
    &    GW170202 & 1170079035.72 &       $4.19$ \\
        \tabline
    \end{tabular}
    \caption{Inverse false-alarm rate for the events considered in this work assigned by our primary search \cite{pipeline, BBH_O2} with an improved detection statistic. Note that GW170817A was found in a different, targeted search \cite{fishing}. The IFAR reported here is referred to the individual observing run and template bank each event was found in, without penalizing for the fact that we searched two observing runs and five template banks, see \S{\ref{ssec:vt}}. These are all events we found with Hanford--Livingston coincident triggers with an $\rm IFAR>3$ runs in their bank.}
    \label{tab:ifar}
\end{table}

\subsection{Search completeness}

A binary merger might fail to be detected for a variety of reasons. In our primary search \cite{pipeline}, a successful detection requires identifying triggers in coincidence at both LIGO detectors, and these coincident triggers must survive a battery of signal quality tests (vetoes). Finally, the candidate has to stand out from the noise background in order to be detected with any significance.

\begin{figure*}
    \centering
    \subfloat[Injections in O1 (top) and O2 (bottom). \label{fig:recovery_fraction}]{%
        \includegraphics[width=.5\linewidth]{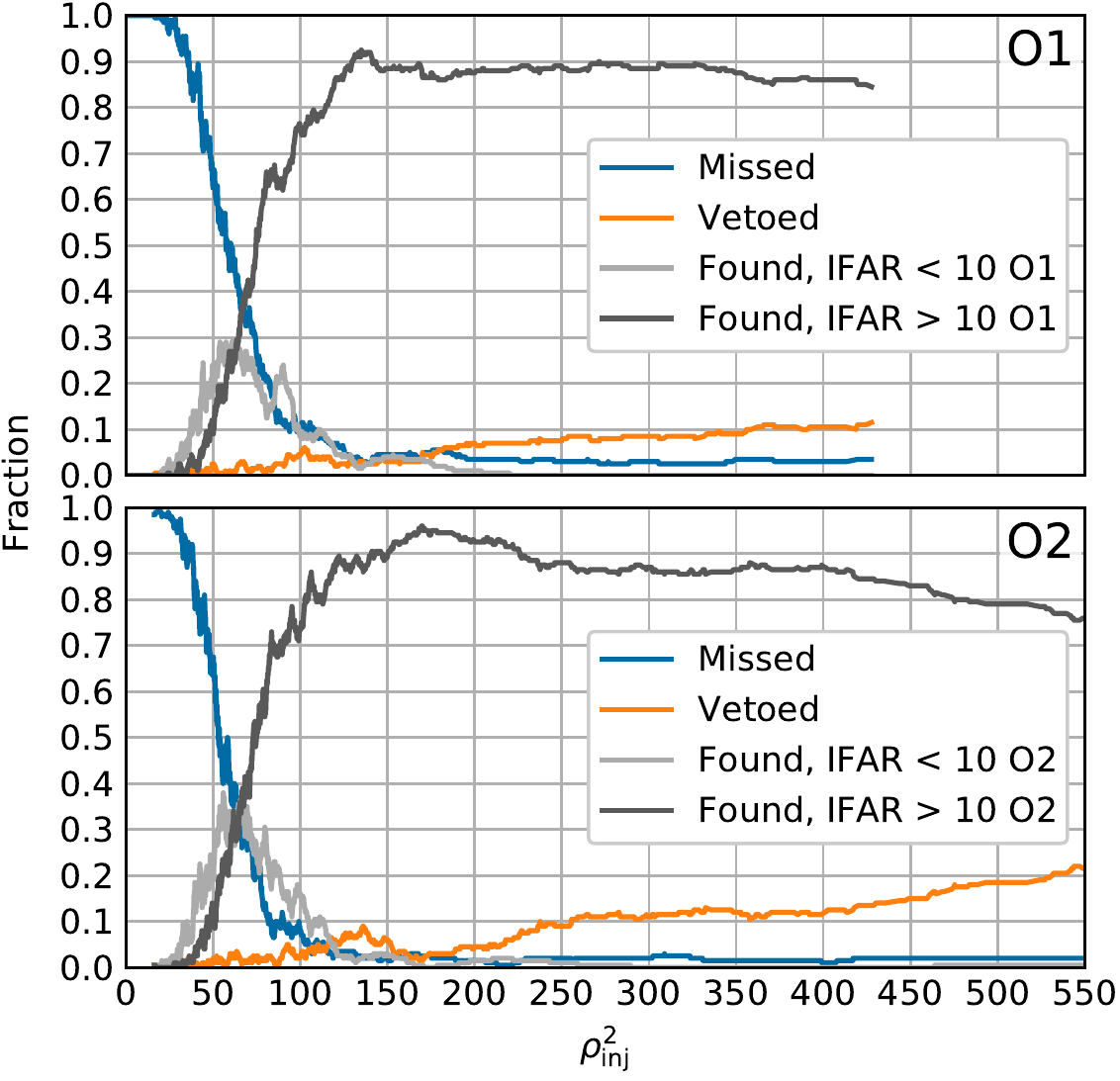}}%
    \subfloat[Injections in O2. \label{fig:inj_scatter}]{%
        \includegraphics[width=.5\linewidth]{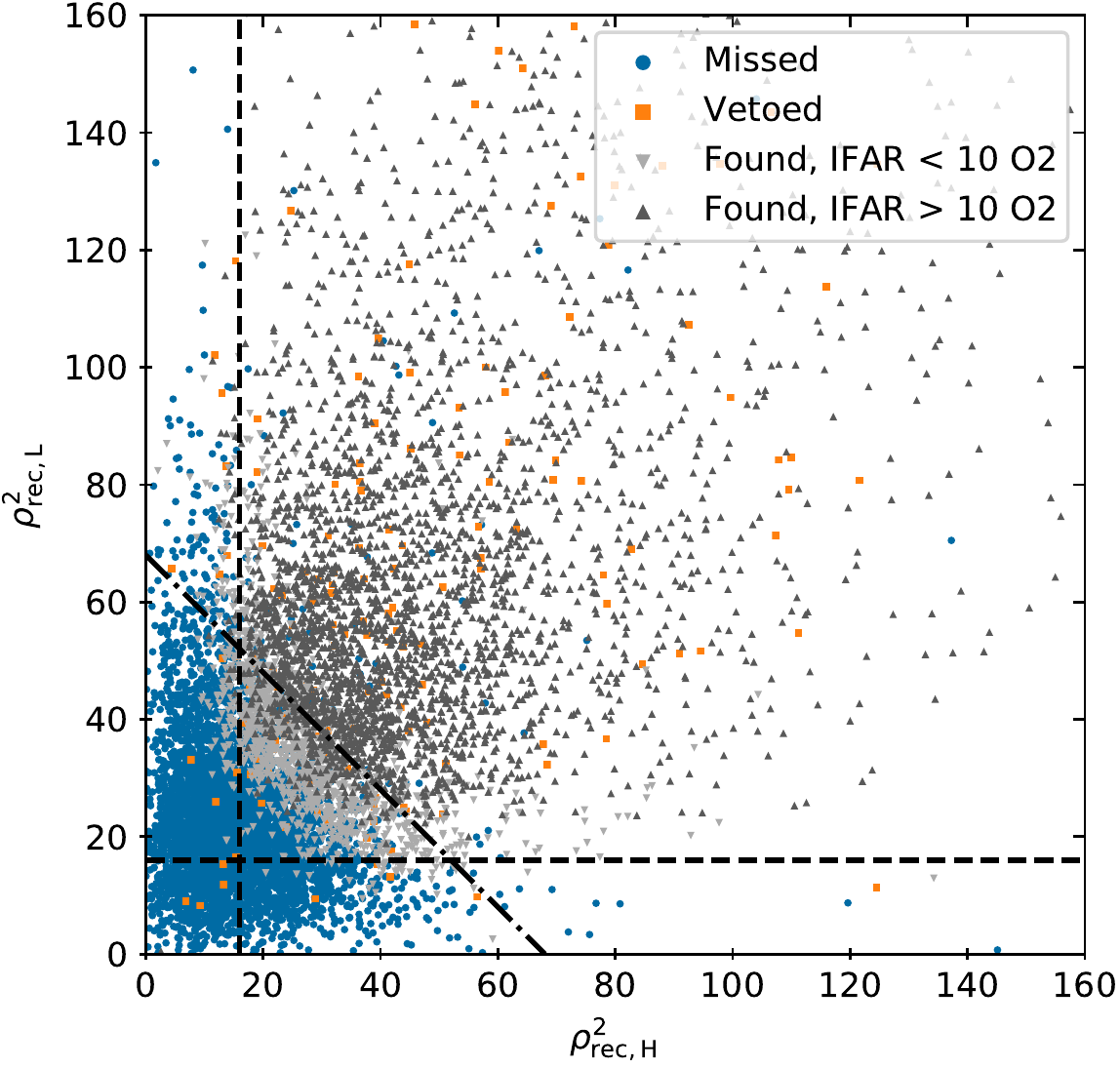}}
    \caption{Probabilities for the possible outcomes of an injection as a function of SNR in our primary search for triggers in Hanford--Livingston coincidence. Only injections with parameters within the target region of our template bank \cite{templatebank_paper}, and that happened during times flagged by both the LVC \cite{GWOSC} and our pipeline \cite{pipeline} as valid for search are included. The four outcomes are exhaustive.
    \textit{Left panel:} Outcome probabilities as a function of injected squared SNR, for the O1 (top) and O2 (bottom) observing runs.
    \textit{Right panel:} Outcomes in the O2 observing run in terms of recoverable squared SNRs at the Hanford and Livingston detectors. Dashed lines are drawn at the single-detector collection threshold $\rho_{{\rm rec \, H, L}}^2 = 16$ and at the approximate Gaussian limit $\rho^2_{\rm rec} = 68$. Several injections with $\rho_{\rm rec}^2 > 68$ have $\rho_{\rm rec, H}^2 < 16$ due to disparate detector responses \citep{fishing}.}
\end{figure*}

As a diagnosis of the performance of our search pipeline, in Fig.~\ref{fig:recovery_fraction} we show the recovery fraction of injections as a function of their squared injected SNR (defined below), as well as the relative frequency of various failure modes.
We include in Fig.~\ref{fig:recovery_fraction} only injections with parameters within the target region of our template bank \cite{templatebank_paper}, and that happened during times flagged by both the LVC \cite{GWOSC} and our pipeline \cite{pipeline} as valid for search.
Injections labeled `Missed' (blue) are those that failed to produce a coincident trigger. 
These include cases when the recovered signal in one of the two LIGO detectors was below our collection threshold ($\rho^2 < 16$), or when a noise transient caused a different template to generate a louder trigger in one of the detectors (some of these second cases could have passed a full `production' search, since we did not do data-cleaning as part of this injection campaign). 
`Vetoed' injections (orange) are those that triggered the signal quality checks, and consequently rejected. We designed these tests to have a false positive rate of a few percent with Gaussian noise for signals having $\rho < 20$ (with an increased rate for much louder events). This is in line with what we observe in Fig.~\ref{fig:recovery_fraction}. Finally, we distinguish injections that we found below or above a moderate inverse false-alarm rate of 10 observing runs in their template bank (of which we searched five). We show these in light and dark gray, respectively; the four curves in Fig.~\ref{fig:recovery_fraction} add up to unity. We derived the fractions using a sliding window in $\rho_{\rm inj}^2$ that averages over 200 contiguous injection samples. 

Since we performed independent searches for five different template banks, the same injection might have a different outcome in each search; and likewise at each of the two LIGO detectors. We summarized this information by assigning to each injection the latest stage it got to in the worst detector, and (then) the best bank. 

We distinguish between `injected' and `recoverable' squared SNR; respectively:
\begin{align}
    \rho_{\rm inj}^2 &= \sum_{k\in{\rm H, L}} \langle h \mid h\rangle_k, \\
    \rho_{\rm rec}^2 &= \sum_{k\in{\rm H, L}} \langle d \mid h\rangle_k,
\end{align}
where $d$ and $h$ are the strains of the data and injection respectively.
Both $\rho_{\rm inj}^2$ and $\rho_{\rm rec}^2$ depend on the signal parameters as well as the detector sensitivities and orientations at the time of the event. Note that $\rho_{\rm inj}^2$ is independent of the particular noise realization and thus unobservable, unlike $\rho_{\rm rec}^2$. Both are independent of the template bank of the search.

We find that the results for O1 and O2 are comparable. The completeness of the search saturates around 90--95\% for signals with $\rho^2_{\rm inj} \gtrsim 150$. At lower SNR values, the dominant failure mode is missed injections, and at higher values it is false rejections (vetoes).

Figure~\ref{fig:inj_scatter} shows the outcome of the injections made in the O2 run (with the same cuts used for Fig.~\ref{fig:recovery_fraction}) scatter-plotted in the $\rho_{\rm rec, L}^2, \rho_{\rm rec, H}^2$ plane. Recall that by virtue of Eq.~\eqref{eq:pinj} the injection distribution approximately follows that of astrophysical events in the high SNR limit $\rho^2 > \rho^2_{\rm th} = 60$.
The incoherent detection limit of our primary search can be approximated by a Gaussian noise limit $\rho^2_{\rm rec} > 68$ and single-detector collection thresholds $\rho^2_{\rm rec\, H, L} > 16$ \cite[figure 6a]{BBH_O2}.
60\% of the injections with $\rho^2_{\rm rec} > 68$ that are missed from the primary search (or 4.6\% of all injections with $\rho^2_{\rm rec} > 68$) have $\rho^2_{\rm rec, H} < 16$, i.e. below our single-detector collection threshold, even though their network SNR is above the Gaussian limit for detection. We performed a targeted search for such signals in \cite{fishing}, which for computational limitation we do not reproduce here with injections. Such events would stand a second chance of being found in the targeted search, which is not accounted for in Fig.~\ref{fig:recovery_fraction}. Figure~\ref{fig:inj_scatter} also supports the approximate incoherent detection thresholds for our pipeline used in \cite[figure 6a]{BBH_O2}.

\subsection{Sensitive volume-time} \label{ssec:vt}

\begin{figure}
    \centering
    \includegraphics[width=\linewidth]{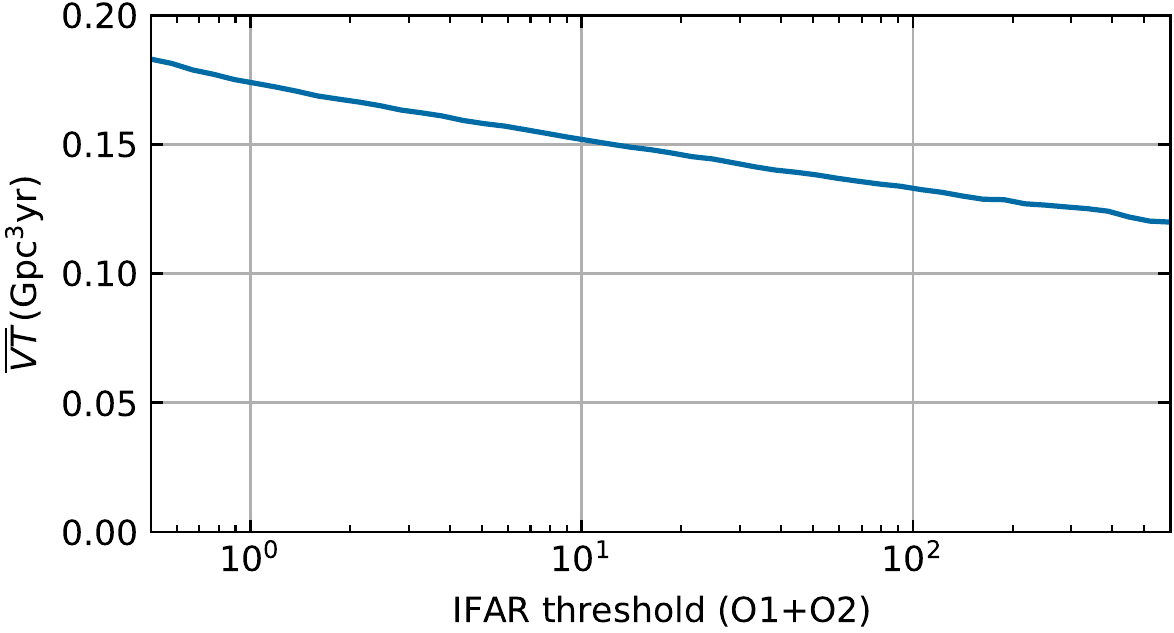}
    \caption{Sensitive volume--time of our search pipeline on O1 and O2 as a function of IFAR threshold for the default population model defined in Eq.~\eqref{eq:fhat}. The IFAR is referred to the full O1+O2 search and accounts for the fact that multiple template banks were searched per Eq.~\eqref{eq:FAR}.}
    \label{fig:VT}
\end{figure}

Quantifying the sensitive volume-time requires defining a detection threshold, see Eq.~\eqref{eq:g_est}. We will define this threshold in terms of IFAR, that can be compared across different search pipelines.
We measure FAR empirically using timeslides, i.e. adding artificial time-shifts between the Hanford and Livingston data streams to generate background triggers, and counting the number of background triggers that have a better detection statistic than a trigger of interest. Our search procedure divides the BBH parameter space into 5 template banks that are explored independently except for the restriction that any trigger is assigned to only one search; further, the O1 and O2 observing runs are analyzed separately.
Therefore, the FAR we obtain has units of per bank per $\overline{VT}$ of the relevant observing run.\footnote{This is the reason we report FARs in units of observing runs instead of physical time---the ranking statistic includes a time-dependent volumetric correction factor to account for the significant and systematic changes in the network sensitivity over the run \cite{pipeline}. If the network sensitivity were constant during the observing run, the units ${\rm ``O1"} \approx \SI{46}{days}$ and ${\rm ``O2"} \approx 118$ days.}
To aid eventual comparisons we express them per all our O1 and O2 BBH searches combined, for which we use $\rm (O1+O2)$ as notation.
For a trigger $j$:
\begin{equation} \label{eq:FAR}
    \begin{split}
        {\rm FAR}_j
        &= N_b(\tilde \rho > \tilde \rho_j)[{\rm run}_j \,{\rm bank}_j]^{-1} \\
        &= N_b(\tilde \rho > \tilde \rho_j)\frac{(\overline{VT})_{\rm O1+O2}}{(\overline{VT})_{{\rm run} j}} N_{\rm banks} [\rm O1 + O2]^{-1},
    \end{split}
\end{equation}
where $\tilde \rho$ is the detection statistic of our pipeline and $N_b(\tilde \rho > \tilde \rho_j)$ is the expected number of background triggers above $j$ estimated from timeslides. We use Eq.~\eqref{eq:FAR} to aggregate the results from all BBH banks and observing runs.
Note that the $\overline{VT}$ estimation in Eq.~\eqref{eq:g_est} requires a threshold on the FAR, and conversely the FAR in Eq.~\eqref{eq:FAR} necessitates a computed $\overline{VT}$ ratio between runs. We find a self-consistent solution numerically. Using the default population model defined in Eq.~\eqref{eq:fhat} and an IFAR threshold of $1(\rm O1+O2)$ we obtain $\overline{VT}_{\rm O2} / \overline{VT}_{\rm O1} = 4.13$. This is approximately valid for other population models and thresholds as well since the dependence largely cancels in the ratio. We show the result for $\overline{VT}$ as a function of the IFAR threshold in Fig.~\ref{fig:VT}, for the default population model of Eq.~\eqref{eq:fhat}.

\section{Computation of the reference \texorpdfstring{$\pastro$}{p-astro}}
\label{app:pastro}

As per Eq.~\eqref{eq:pastro}, $\pastroi$ depends on the ratio of the trigger densities expected from foreground and background locally at the data $d_i$ of each trigger. In principle, $d_i$ consists of the strain time series, or equivalently its Fourier modes. In practice we do not have a reliable model for the background trigger density in terms of these, as we do for the astrophysical events (Eqs.~\eqref{eq:dNa/dd} and \eqref{eq:Gaussian_like}). Thus, we instead approximate (for all events except GW170608 and GW170817A, see below):
\begin{equation}\label{eq:dNa/dNb}
    \diff{N_a}{N_b} \bigg|_{d_i}(\lambda_0)
    \approx \frac{\rmd N_a / \rmd\tilde\rho^2}{\rmd N_b / \rmd\tilde\rho^2}
        (\tilde \rho^2_i \mid \mathcal T \approx \mathcal T_i, \lambda_0),
\end{equation}
where $\tilde \rho^2$ is the detection statistic of our pipeline \cite{pipeline} and $\mathcal T$ identifies the waveform template associated to a trigger.
We expect these variables to contain most of the discerning power between astrophysical and background events. The detection statistic is an estimator of the likelihood ratio between the astrophysical and noise hypotheses that incorporates information from the trigger SNR at Hanford and Livingston, the phase and time differences between detectors and the instantaneous detector sensitivities. It includes only very coarse information about the signal parameters, based on which subbank the triggering template resides in.
Since both the astrophysical and background trigger rate densities exhibit dependence on the triggering template, we incorporate the template identity in Eq.~\eqref{eq:dNa/dNb}.

We compute Eq.~\eqref{eq:dNa/dNb} for each trigger as follows.
We first generate large sets of astrophysical and background triggers. For the former we use the injections described in \S\ref{ssec:distribution_choice} and for the latter we use the method of timeslides.
We restrict to the triggers that have a similar template to the triggering template $\mathcal T_i$: we demand that they are found in the same chirp mass bank \cite{templatebank_paper} and have a match $\langle \mathcal T \mid \mathcal T_i \rangle$ above some threshold. We choose the highest match threshold that admits at least 100 injection and 100 background triggers.\footnote{For bank \texttt{BBH 0} \cite{templatebank_paper} we only recovered 24 injections in O2; we just include those in the computation of the $\pastro$ of 170412B.} This achieves a compromise between making a density measurement that is local in intrinsic-parameter space and that has an acceptable statistical error. We then do a kernel density estimation of $\rmd N_a / \rmd\tilde\rho^2$ and $\rmd N_b / \rmd\tilde\rho^2$, using these triggers with weights (see Eq.~\eqref{eq:pinj})
\begin{align}
    w_j^{\rm inj} &= \frac{R_0\,Z}{N_{\rm inj} \, \hat p_{\rm det}(\theta_j)}
        \label{eq:w_inj}\\
    w^{\rm bg} &= \frac{1}{N_{\rm timeslides}}
\end{align}
for injections and background respectively.

As mentioned in Appendix~\ref{app:sensitivity}, for GW170818 we obtained a $\pastro(\lambda_0)=0.92$ based on Hanford and Livingston data, higher than reported in \cite{BBH_O2} owing to an improved version of the coherent score, as we described in Sec.~\ref{sec:inj_campaign}. Note the GstLAL pipeline \cite{gstlal} found $\pastro=1$ by including Virgo data. As discussed in \S\ref{ssec:likelihood_evaluation}, following \citet{GW170608} we set $\pastro=1$ for GW170608.

The event GW170817A was found in a targeted search for signals that are loud in the Livingston detector and faint in the Hanford detector \cite{fishing}, as opposed to our primary search for signals in Hanford--Livingston coincidence \cite{pipeline, BBH_O2}. Most of its significance comes from being the loudest Livingston trigger apart from previously confirmed confident signals. As such, the method of timeslides cannot be used to generate empirical background for this event. We compute its $\pastro(\lambda_0)$ following \cite[eqs.~(5) and (6)]{fishing}: we define the clean region of parameter space as those templates with chirp mass $\mathcal M > \SI{10}{M_\odot}$ for which there were $\leq 5$ loud Livingston triggers ($\rhol^2 > 60$) in O2 from similar templates (match $>0.9$) in times where the Hanford detector was also operating. We obtain the expected number of triggers in this region with $\rhol^2 > 66$ by counting the injections that satisfy all these conditions weighted per Eq.~\eqref{eq:w_inj}. We set the expected number of background events to 1.

Recently, \citet{Ashton2020} have concluded that the event GW151216 \cite{GW151216} has $\pastro = 0.03$, based on an analysis of background triggers obtained with a different pipeline, as well as foreground triggers generated under an astrophysical prior isotropic in spin directions. A similar method was used by \citet{Pratten2020}. Those analyses overlook the fact that different pipelines treat the systematics in the data differently and thus suffer from different backgrounds. As an example, our pipeline applies different data quality checks and signal consistency vetoes. Even within our pipeline, removing or modifying these stages would significantly lower the $\pastro$ of near-threshold triggers like GW151216. These tests are not applied in the analysis of \cite{Ashton2020, Pratten2020}. They instead characterize the background in terms of its projection onto parameter space when modeled as a GW signal in Gaussian noise. That is a different test that a priori does not have the same discerning power between signals and noise. Ultimately, the choices that maximize the pipeline sensitivity (see Appendix~\ref{app:sensitivity}) should be pursued. As a result of these considerations, $\pastro$ is inherently a pipeline-dependent quantity.
In addition, $\pastro$ depends on the astrophysical population model. As shown in Table~\ref{tab:pastro}, for GW151216 (as well as other near-threshold events with nonzero spin) $\pastro$ is particularly sensitive to the spin distribution. \citet{Ashton2020, Pratten2020} used an isotropic spin model, under which the astrophysical interpretation is indeed strongly disfavored.
We emphasize however that for different spin models GW151216 has a sizable $\pastro$. We should consistently account for this dependence when interpreting GW151216 and assessing its implications for the astrophysical BBH population.

\section{Differences with \texorpdfstring{\citet{Galaudage2020}}{Gaulaudage et al (2019)}}
\label{app:differences}

\citet{Galaudage2020} have presented a framework similar but inequivalent to the one presented in Section~\ref{sec:framework}. In this appendix we compare both treatments and identify the differences between them. Our point of comparison is the model likelihood, our Eq.~\eqref{eq:like_ratio} or their equation~(36), which are not compatible.

The correspondence between their notation and ours is as follows. Their astrophysical hypothesis prior is
\begin{equation} \label{eq:xi}
     \xi \equiv \frac{N_a}{N_a + N_b}
 \end{equation} 
in our notation. Their signal likelihood is 
\begin{multline} \label{eq:signal_like}
        \mathcal L (d \mid \Lambda, {\rm det})
        = \frac{\mathcal V_{\rm tot}}{\mathcal V(\Lambda)}
            \mathcal L(d \mid \Lambda) \\
        \equiv \frac{1}{N_a} \diff{N_a}{d}
        = \frac{1}{\overline{VT}(\lambda')}
            \int \rmd \theta f(\theta \mid \lambda') P(d \mid \theta),
\end{multline}
where their $\Lambda \equiv \lambda'$, $\mathcal V(\Lambda) \equiv \overline{VT}(\lambda')$ and we have used Eqs.~\eqref{eq:dNa/dd}, \eqref{eq:R*f} and \eqref{eq:R*g}.
Their noise likelihood is
\begin{equation} \label{eq:noise_like}
    \begin{split}
        \mathcal L (d \mid \varnothing, {\rm det})
            &\equiv \frac{1}{N_b} \diff{N_b}{d} \\
            &= \frac{N_a}{N_b} \cdot \diff{N_b}{N_a}
                \cdot \frac{1}{N_a} \diff{N_a}{d} \\
            &= \frac{\xi}{1 - \xi} \cdot \frac{1 - \pastro}{\pastro}
                \cdot \mathcal L (d \mid \Lambda, {\rm det}),
    \end{split}
\end{equation}
using Eqs.~\eqref{eq:pastro}, \eqref{eq:xi} and \eqref{eq:signal_like}. Equation~\eqref{eq:noise_like} is \cite[eq.~(7)]{Galaudage2020} generalized to account for selection effects.

In their equation (29), \citet{Galaudage2020} use a prescription for the normalization of the background term in the likelihood, $p_\varnothing = \mathcal V(\Lambda_0) / \mathcal V_{\rm tot}$, which holds for $\Lambda$ on the vicinity of $\Lambda_0$. Instead, the exact expression that satisfies Eq.~\eqref{eq:noise_like} is
\begin{equation} \label{eq:selection_correction}
    p_\varnothing(\Lambda) = \frac{\mathcal V(\Lambda)}{\mathcal V_{\rm tot}}.
\end{equation}

In \cite[eq.~(35)]{Galaudage2020} it is stated that $(p_{\rm astro}^{-1}-1) \mathcal L(d \mid \Lambda)$ is independent of $\Lambda$, so they equate this term to its value for a fiducial $\Lambda_0$. However, $\pastro$ also depends on the rate of mergers, so the fiducial model should include the rate (encoded in $\xi$). Additionally, their equation~(35) relies on equation~(7) which does not include selection effects. The correct form of \cite[eq.~(35)]{Galaudage2020} is thus
\begin{multline} \label{eq:noise_like_correction}
    \frac{\xi}{1 - \xi}
        \frac{1 - \pastro(\xi, \Lambda)}{\pastro(\xi, \Lambda)}
        \frac{\mathcal L(d \mid \Lambda)}{\mathcal V(\Lambda)} \\
    = \frac{\xi_0}{1 - \xi_0}
        \frac{1 - \pastro(\xi_0, \Lambda_0)}{\pastro(\xi_0, \Lambda_0)}
        \frac{\mathcal L(d \mid \Lambda_0)}{\mathcal V(\Lambda_0)},
\end{multline}
where $\xi_0, \Lambda_0$ are the values used to compute the fiducial $\pastro$.

Lastly, in \cite[eq.~(34)]{Galaudage2020} the expression $\xi = N/n$ is used, where $N \equiv \ntrig$ is the number of triggers and $n$ the number of data segments analyzed. But $\xi$ encodes a (model dependent) prior expectation of the ratio of counts of astrophysical and noise events, and not the actual outcome of the experiment. The correct expression is instead
\begin{equation} \label{eq:xi_correction}
    \xi = \frac{R \mathcal V(\Lambda)}{R \mathcal V(\Lambda) + R_g T_{\rm obs}}.
\end{equation}

Once the changes in Eqs.~\eqref{eq:selection_correction}, \eqref{eq:noise_like_correction} and \eqref{eq:xi_correction} have been applied, their equation (36) becomes
\begin{widetext}
    \begin{equation} \label{eq:equivalence}
        \begin{split}
            P(\{d_i\}, N \mid \Lambda) 
            &= \frac{e^{-(R \mathcal V(\Lambda) + R_g T_{\rm obs})}
                (R \mathcal V(\Lambda))^N}{N!}
                \prod_i \frac{\mathcal V_{\rm tot}}{\mathcal V(\Lambda)}
                \frac{\mathcal L(d_i \mid \Lambda)}{\pastroi(R, \Lambda)} \\
            &= \frac{e^{-(R \mathcal V(\Lambda) + R_g T_{\rm obs})}
                (R_0 \mathcal V_{\rm tot})^N}{N!}
                \prod_i \left(
                \frac{1 - \pastroi(R_0, \Lambda_0)}{\pastroi(R_0, \Lambda_0)}
                \frac{R_g}{R_{g, 0}} + \frac{R}{R_0}
                \frac{\mathcal L(d_i \mid \Lambda)}{\mathcal L(d_i \mid \Lambda_0)}
                \right) \mathcal L(d_i \mid \Lambda_0),
        \end{split}
    \end{equation}
\end{widetext}
which is compatible with Eq.~\eqref{eq:like_ratio}. The rate $R$ does not factor out, so we cannot marginalize it analytically in Eq.~\eqref{eq:equivalence} without expanding the binomial first.

As a consequence of Eq.~\eqref{eq:noise_like_correction},
\begin{equation}
    \pastro(\xi, \Lambda) = \left(
    \frac{\xi_0}{1- \xi_0}
    \frac{\mathcal L(d \mid \Lambda_0)}{\mathcal V(\Lambda_0)}
    \frac{1- \xi}{\xi}
    \frac{\mathcal V(\Lambda)}{\mathcal L(d \mid \Lambda)} + 1 \right)^{-1},
\end{equation}
instead of \cite[eq.~(42)]{Galaudage2020}. This difference might explain why Table~\ref{tab:pastro} does not reproduce the results of \cite[table~II]{Galaudage2020}.
We emphasize that the fiducial $\pastro$ have to correspond to $R_0, \Lambda_0$ and reported values must be interpreted with this in mind. In Table~\ref{tab:pastro} we have computed the $\pastro$ of the top triggers our pipeline found under a specific astrophysical model $\lambda_0$ to facilitate this task.

\section{Robustness of the reweighting procedure}
\label{app:robustness}

Our framework involves Monte Carlo computation of integrals by reweighting samples: the numerator and denominator of Eq.~\eqref{eq:wi_est} use source parameter estimation samples, and Eq.~\eqref{eq:g_est}) uses injection samples. These methods are subject to stochastic errors, especially if the target and proposal distributions are mismatched. We have chosen the proposal distributions with this consideration in mind, in this appendix we show that our procedure indeed achieved sufficient robustness.

\begin{figure*}
     \includegraphics[width=.37\linewidth, valign=t]{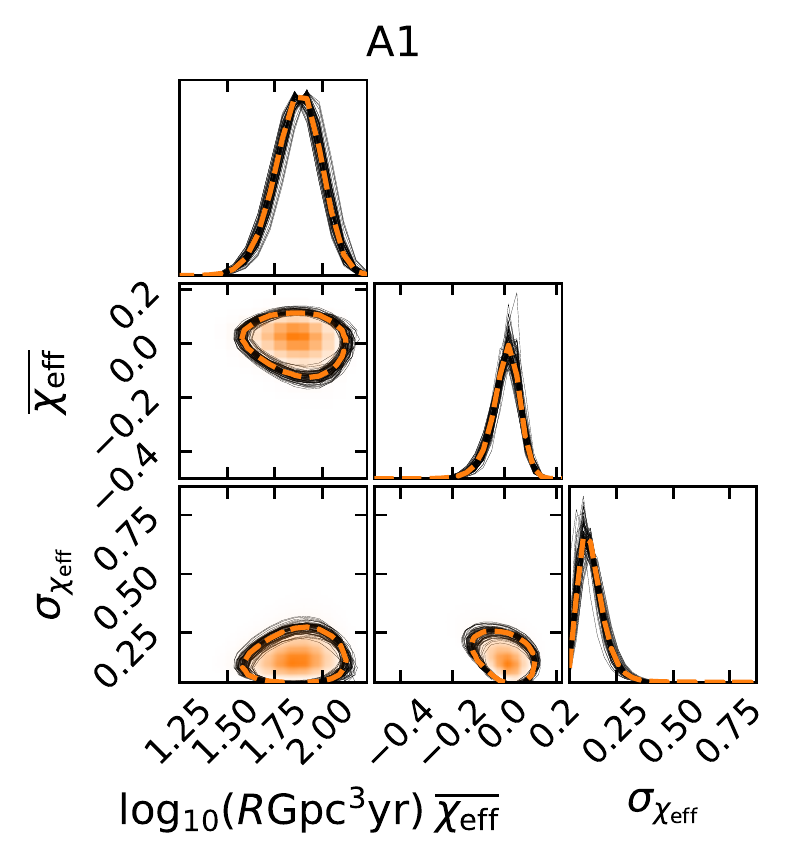}%
     \includegraphics[width=.37\linewidth, valign=t]{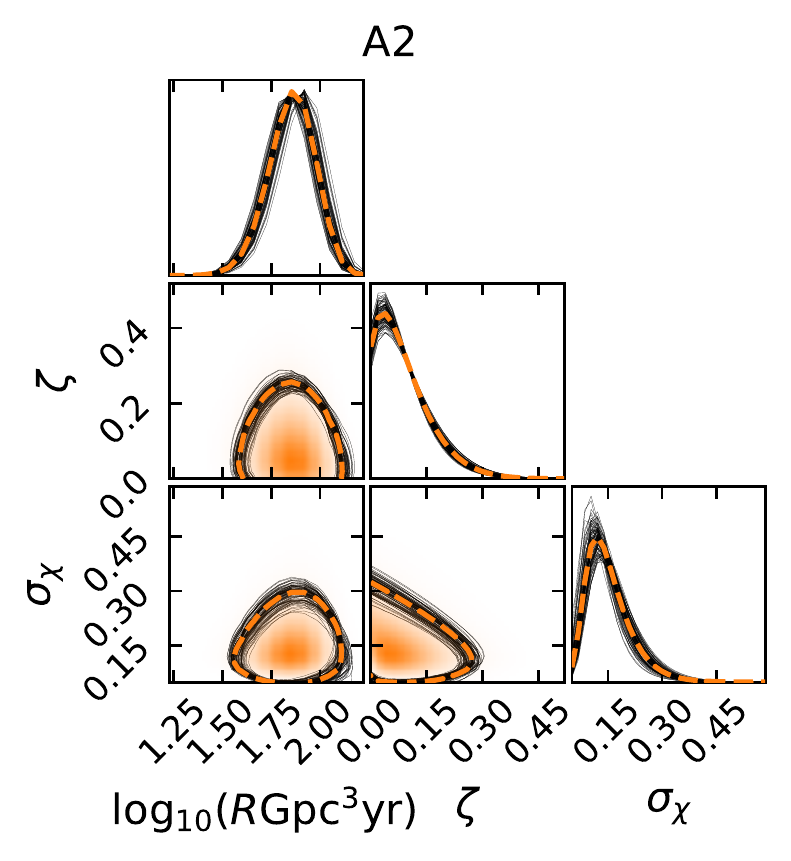}%
     \includegraphics[width=.26\linewidth, valign=t]{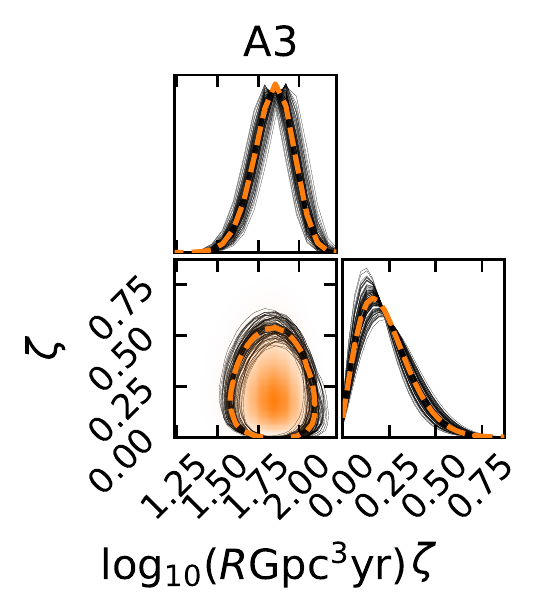}%
     \\
    \includegraphics[width=.49\linewidth, valign=t]{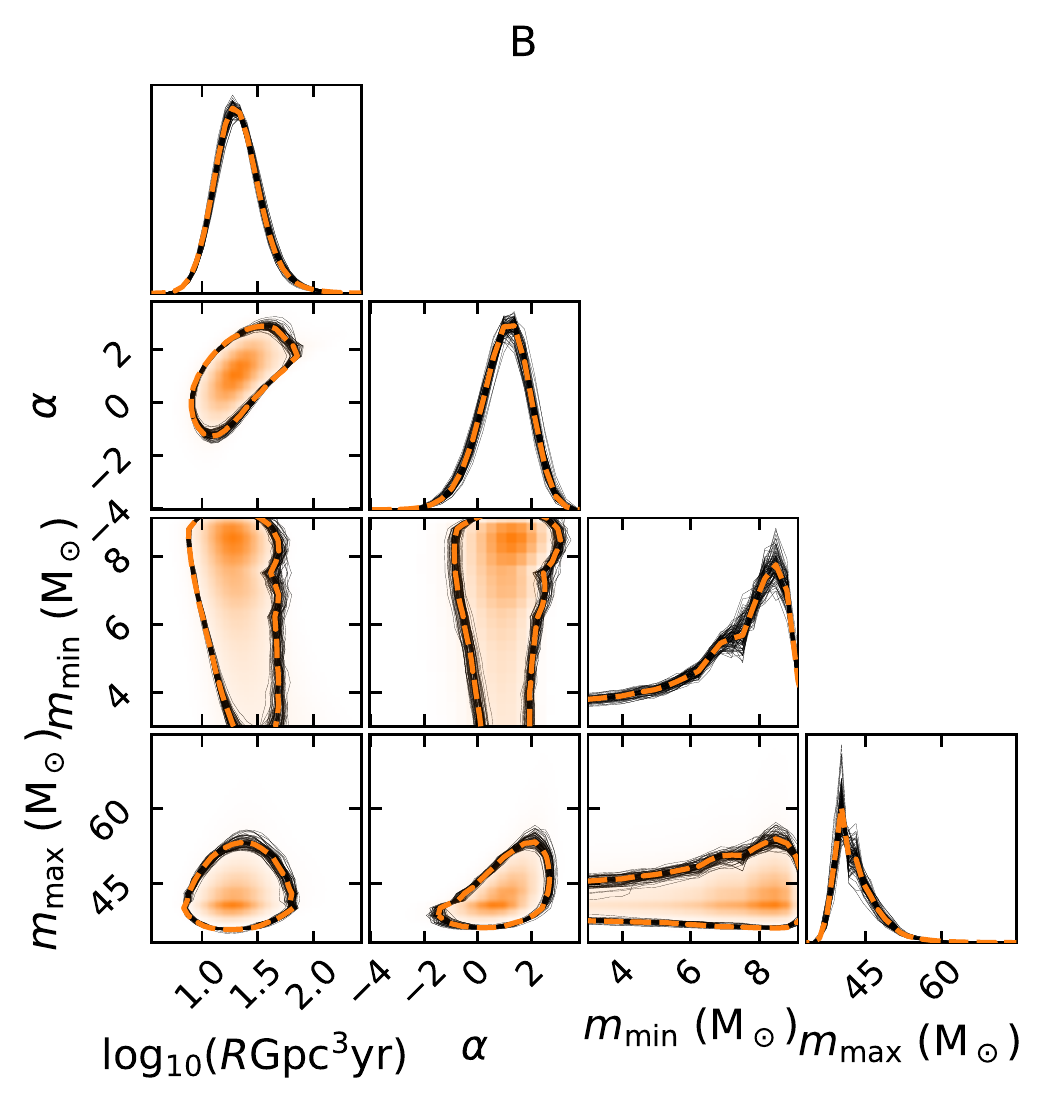}%
    \hspace{-.0301\linewidth}%
    \includegraphics[width=.27\linewidth, valign=t]{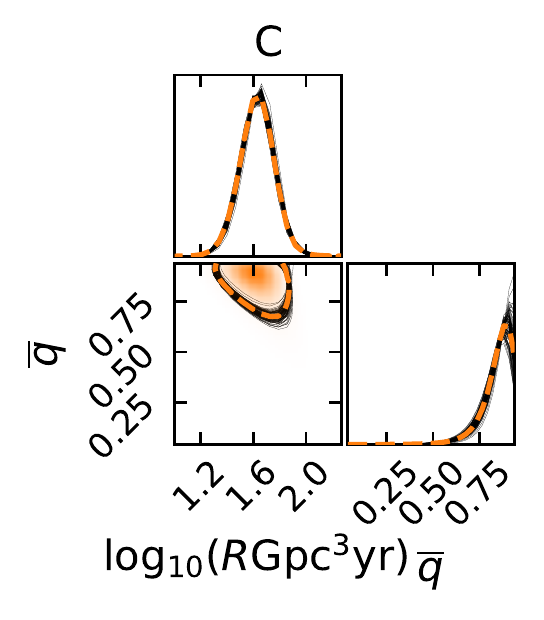}%
    \includegraphics[width=.27\linewidth, valign=t]{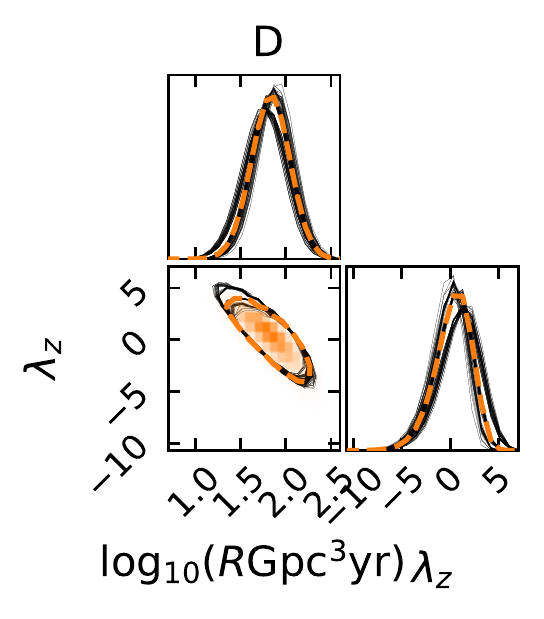}%
    \caption{100 bootstrap realizations for the population Models~\ref{model:gaussian_chieff}--\ref{model:z_powerlaw} explored in this work (thin black), along with our results from \S\ref{ssec:spin}--\ref{ssec:redshift} (dashed orange). These quantify the impact of stochastic errors associated with the various Monte Carlo integrations involved, which remains small at the current uncertainty level. Contours enclose 90\% of the distribution.}
    \label{fig:robustness}
\end{figure*}

We use the bootstrap method to estimate the impact that stochastic error in these integrations has on the population inference. For each population model, we do 100 bootstrap repetitions of the inference of population parameters $\lambda$, each time employing a set of samples taken randomly with replacement from the corresponding original sets for all Monte Carlo integrations involved---i.e., from the source parameter estimation samples of each event for the computation of $\mathcal W_i(\lambda' \mid \lambda_0')$ in Eq.~\eqref{eq:wi_est}, and from the set of injections for $\overline{\mathcal{VT}}(\lambda')$ in Eq.~\eqref{eq:g_est}. The collection of these inference realizations may be interpreted as the ensemble of solutions compatible with our stochastic sampling errors.
Figure~\ref{fig:robustness} shows the results of this exercise for Models~\ref{model:gaussian_chieff}--\ref{model:z_powerlaw} and the combined GWTC-1 and IAS catalog. Thin black lines show the 100 bootstrap realizations and dashed orange lines show our original results. These stochastic errors are smaller than the uncertainty levels, providing evidence that the proposal distribution choices and the number of injections and parameter estimation samples were adequate to achieve a robust reweighting.

\bibliographystyle{apsrev4-1-etal}
\bibliography{main}

\end{document}